\documentclass[aoas,preprint]{imsart}
\usepackage{color}
\usepackage{float}
\usepackage{amsmath}
\usepackage{amssymb}
\usepackage{graphicx}
\usepackage{esint}
\usepackage{natbib}
\usepackage{multirow}
\usepackage{url}
\setcitestyle{square}

\arxiv{arXiv:0000.0000}

\startlocaldefs
\endlocaldefs

\begin{document}

\begin{frontmatter}

% "Title of the paper"
\title{Bayesian Hidden Markov Tree Models for Clustering Genes with Shared
Evolutionary History}
\runtitle{Clustering Genes with Shared Evolutionary History}

% indicate corresponding author with \corref{}
%\author{\fnms{John} \snm{Smith}\corref{}\ead[label=e1]{smith@foo.com}\thanksref{t1}}
% \thankstext{t1}{Thanks to somebody} 
% \address{line 1\\ line 2\\ printead{e1}}
% \affiliation{Some University}

\begin{aug}
\author{\fnms{Yang} \snm{Li}\thanksref{m1,m4,a1}\ead[label=e1]{yangli.stat@gmail.com}},
\author{\fnms{Shaoyang} \snm{Ning}\thanksref{m1,a1}\ead[label=e2]{shaoyangning@fas.harvard.edu}},
\author{\fnms{Sarah E.} \snm{Calvo}\thanksref{m2,m3,m4},
\ead[label=e3]{scalvo@broadinstitute.org}
\ead[label=u1,url]{http://www.foo.com}}\\
\author{\fnms{Vamsi K.} \snm{Mootha}\thanksref{m5,m2,m3,m4}
\ead[label=e4]{vamsi@hms.harvard.edu}}
\and
\author{\fnms{Jun S.} \snm{Liu}\thanksref{m1}
 \ead[label=e5]{jliu@stat.harvard.edu}}

\runauthor{Y. Li et al.}
\thankstext{a1}{These authors contributed equally to the manuscript.} 
\affiliation{Harvard University\thanksmark{m1}, Broad
Institute\thanksmark{m2}, Harvard Medical School\thanksmark{m3}, Massachusetts General Hospital\thanksmark{m4}, and Howard Hughes Medical Institute\thanksmark{m5}}

\address{Y.Li\\Department of Statistics\\
Harvard University\\Cambridge, MA 02138
\\USA\\
\printead{e1}\\
}

\address{S.Ning\\Department of Statistics\\
Harvard University\\Cambridge, MA 02138
\\USA\\
\printead{e2}\\
}

\address{S.E.Calvo\\Broad
Institute\\ Cambridge, MA 02142\\USA\\
\printead{e3}}

\address{V.K.Mootha\\
Howard Hughes Medical Institute\\
Department of Systems Biology\\ Harvard Medical School\\
Boston, MA 02115\\USA\\
Massachusetts General Hospital\\
Boston, MA 02114\\USA\\
\printead{e4}}

\address{J.S.Liu\\Department of Statistics\\
Harvard University\\Cambridge, MA 02138
\\USA\\
\printead{e5}\\
}
\end{aug}

\iffalse
\author{Yang Li, Shaoyang Ning, Sarah E. Calvo, Vamsi K. Mootha, Jun S. Liu\thanks{Yang Li (E-mail: yangli.stat@gmail.com) and Shaoyang Ning (E-mail:
shaoyangning@fas.harvard.edu) are Graduate Students and Jun S. Liu
(E-mail: jliu@stat.harvard.edu) is Professor, Department of Statistics,
Harvard University, Cambridge, MA 02138, USA. Sarah E. Calvo (Email:
scalvo@broadinstitute.org) is Senior Computational Biologist, Broad
Institute, Cambridge, MA 02142. Vamsi K. Mootha (Email: vamsi@hms.harvard.edu)
is Professor, Department of Systems Biology, Harvard Medical School,
Boston, MA 02115 and is Professor, Department of Molecular Biology,
Massachusetts General Hospital, Boston, MA 02114, USA. Vamsi K. Mootha
is a Howard Hughes Medical Investigator.}}
%\author{\fnms{???} \snm{???}\ead[label=e1]{???}}
%\address{\printead{e1}}
%\and
%\author{\fnms{???} \snm{???}\ead[label=e2]{???}}
%\address{\printead{e2}}
%\affiliation{???}
\fi

\begin{abstract}
\textbf{}

Determination of functions for poorly characterized genes is crucial for understanding biological processes and studying human diseases. Functionally associated genes are often gained and lost together through evolution. Therefore identifying co-evolution of genes can predict functional gene-gene associations. We describe here the full statistical model and computational strategies underlying the original algorithm {\it CLustering by Inferred Models of Evolution} (CLIME 1.0) recently reported by us \citep{li2014expansion}. CLIME 1.0 employs a mixture of tree-structured hidden Markov models for gene evolution process, and a Bayesian model-based clustering algorithm to detect gene modules with shared evolutionary histories (termed evolutionary
conserved modules, or ECMs). A Dirichlet process prior was adopted for estimating the number of gene clusters and a Gibbs sampler was developed for posterior sampling. We further developed an extended version, CLIME 1.1, to incorporate the uncertainty on the evolutionary tree structure.  By simulation studies and benchmarks on real data sets, we show that CLIME 1.0 and CLIME 1.1 outperform traditional methods that use simple metrics (e.g., the Hamming distance or Pearson correlation) to measure co-evolution between pairs of genes.

\end{abstract}

%\begin{keyword}[class=MSC]
%\kwd[Primary ]{}
%\kwd{}
%\kwd[; secondary ]{}
%\end{keyword}


\end{frontmatter}

% AOS,AOAS: If there are supplements please fill:
%\begin{supplement}[id=suppA]
%  \sname{Supplement A}
%  \stitle{Title}
%  \slink[doi]{10.1214/00-AOASXXXXSUPP}
%  \sdatatype{.pdf}" 
%  \sdescription{Some text}
%\end{supplement}

\section{Introduction}

The human genome encodes more than 20,000 protein-coding genes, of which a large fraction
do not have annotated function to date \citep{galperin2010complete}.
Predicting unknown member genes to biological pathways/complexes
and the determination of function for  poorly characterized
genes are crucial for understanding biological processes and human
diseases. It has been observed that functionally associated
genes tend to be gained and lost together during evolution \citep{pellegrini1999assigning,kensche2008practical}.
Identifying shared evolutionary history (aka, co-evolution) of genes can help
predict functions for unstudied genes, reveal alternative
functions for genes considered to be well characterized, propose new members of biological pathways, and provide new insights into human diseases.

%In recent years, the availability of diverse genomes has been dramatically increasing - 

\begin{figure}[H]
\begin{centering}
\includegraphics[scale=0.70]{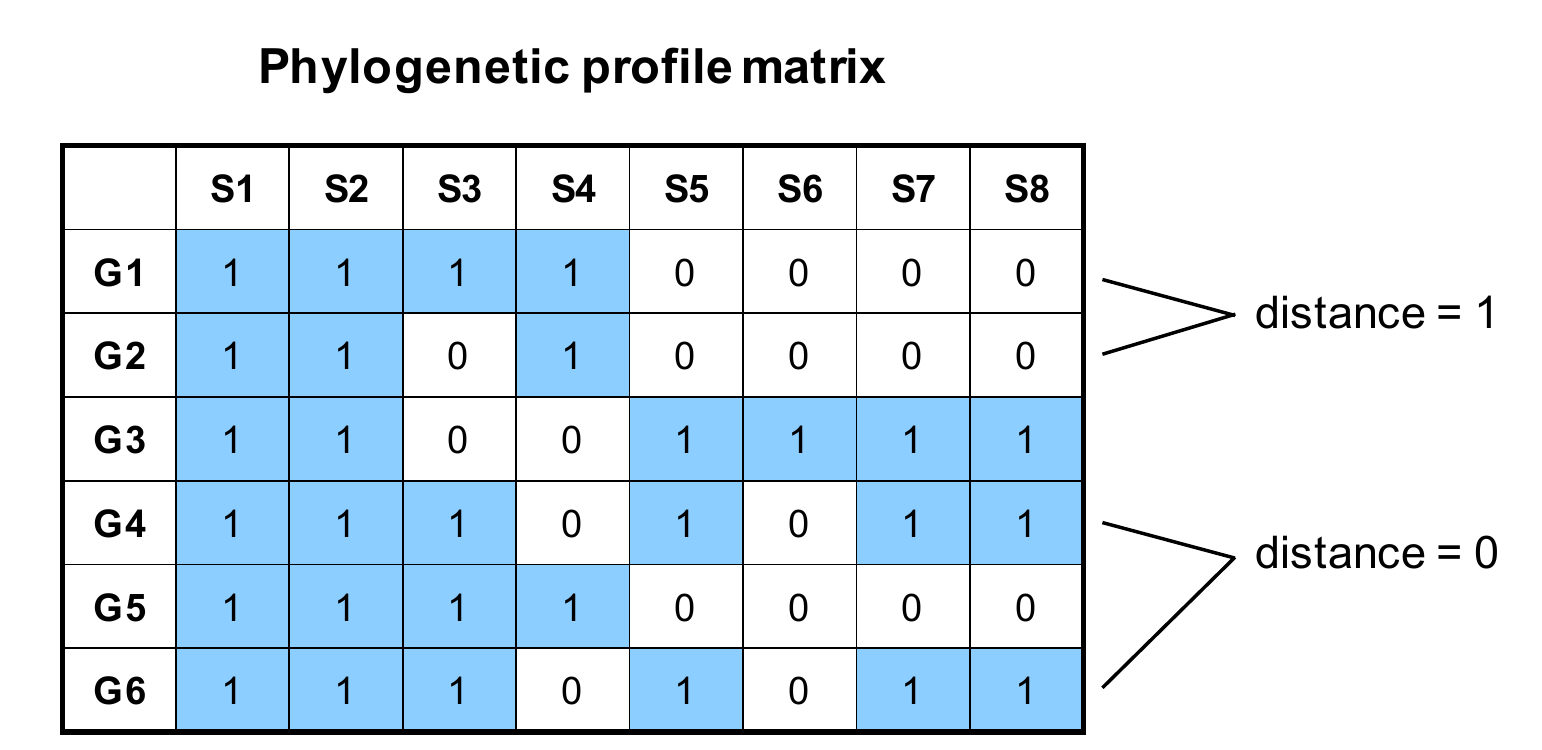}
\par\end{centering}

\caption{A toy example of phylogenetic profile matrix for $N=6$ genes (G1,
..., G6) and $S=8$ species (S1, ..., S8). Blue and white squares
respectively denote presence or absence of genes in corresponding
genomes. G1 and G2 have Hamming distance $1$, while G4 and G6 have
Hamming distance $0$. \label{fig:hamming}}
\end{figure}

The concept of ``phylogenetic profiling'' was first introduced by \cite{pellegrini1999assigning}  to characterize phylogenetic distributions of genes. One can predict a gene's function based on its phylogenetic similarity to those with known functions. %\citet{pellegrini1999assigning} used 
Let the binary phylogenetic profile matrix $\mathbf{X}_{N\times S}$
denote the presence/absence of $N$ genes across $S$ species.  \citet{pellegrini1999assigning} proposed
to measure the ``degree" of co-evolution of a pair or genes $i$ and $j$ as the Hamming distance \citep{hamming1950error} between the $i$th and $j$th rows of $\mathbf{X}$. A toy example
is shown in Figure \ref{fig:hamming}. Various  methods have since been developed (see \citep{kensche2008practical} for
a review) and applied with success in predicting components
for prokaryotic protein complexes \citep{pellegrini1999assigning};
phenotypic traits such as  pili, thermophily, and respiratory tract tropism
\citep{jim2004cross}; cilia \citep{li2004comparative}; mitochondrial
complex I \citep{ogilvie2005molecular,pagliarini2008amitochondrial};
and small RNA pathways \citep{tabach2013identification}. 

Currently there are more than 200 eukaryotic
species with their genomes completely sequenced and about 2,000 species with
full genomes being sequenced (JGI GOLD\footnote{JGI Genome Online Database: https://gold.jgi.doe.gov/}).
The growing availability of genome sequences from diverse species
provides us unprecedented opportunities to chart the evolutionary history
of human genes.  However, existing phylogenetic profiling methods still suffer from some limitations  \citep{kensche2008practical}.
First, most available methods perform only pairwise comparison between an input query gene and a candidate, and are thus unable to discover subtle patterns that show up only after aligning multiple input query genes. Such methods also %lack the potential to take into account the biological pathways, 
cannot handle cases where members in the query gene set  exhibit different phylogenetic profiles. Second, most methods ignore errors in phylogenetic profiles, which are often caused by inaccuracies in genome assembly, gene annotation, and detection of distant homologs \citep{trachana2011orthology}.
Third, most methods (with exceptions of \cite{barker2005predicting,vert2002tree,von2003string,zhou2006inferring}) assume independence across input species, ignoring their 
phylogenetic relationships, e.g., the tree structure of their evolutionary history. These methods are rather sensitive to the organisms' selection  in the analysis. Currently available
tree-based methods, however, are computationally cumbersome and hardly scalable for analyzing large input sets, let alone entire genomes \citep{barker2005predicting,barker2006constrained}. 

To cope with the aforementioned limitations, \cite{li2014expansion} introduced the two-step procedure {\it CLustering by Inferred Models of Evolution} (denoted by CLIME 1.0). %which will be described more comprehensively in this paper. 
In its Partition step, CLIME 1.0 clusters the input gene set $\mathcal{G}$ into disjoint evolutionarily conserved modules (ECMs), simultaneously inferring
the number of ECMs and each gene's ECM membership. In the 
Expansion step, CLIME 1.0 scores and ranks other genes not in  $\mathcal{G}$ according to a log-likelihood-ratio (LLR) statistic for their likelihood of being new members of an inferred ECM. 
%inferred model of evolution compared with the background null model, using a log-likelihood-ratio (LLR) statistic. 
\cite{li2014expansion} systematically applied CLIME 1.0 to over 1,000
human canonical complexes and pathways, resulting in a discovery of  unanticipated co-evolving
components and  new members of important gene sets.

We here provide a full statistical account of CLIME 1.0 and its computational strategies, evaluate CLIME 1.0's performances with extensive simulations, extend it to incorporate uncertainties in the phylogenetic tree structure, and compare CLIME 1.0 with existing methods such as BayesTraits. Finally we apply CLIME 1.0 to gene sets in OMIM (Online Mendelian Inheritance in Man) to reveal new insights on human genetic disorders. Compared with existing methods, by incorporating a coherent statistical model, CLIME 1.0  (1)  takes proper account of the dependency between species; (2) automatically learns the number of distinct evolutionary modules in the input gene set $\mathcal{G}$; (3) leverages information from the entire input gene set to more reliably predict new genes that have arisen with a shared pattern of evolutionary gains and losses; (4) uses the LLR statistic as a principled measure of co-evolution compared to naive metrics (e.g. Hamming distance, Pearson correlation). 

Complementary to the original CLIME 1.0, we further provide an extended version, named CLIME 1.1, which inherits the Bayesian hidden Markov tree model from CLIME 1.0, but further accounts for the uncertainty of the input phylogenetic tree structure by incorporating a prior on the evolutionary tree. Instead of a single, fixed tree as by CLIME 1.0, CLIME 1.1 takes an empirical distribution of tree structures, in addition to the phylogenetic profiles of a given gene set, as input; infers the posterior of the hidden evolutionary histories, hidden cluster (ECM) labels and parameters, as well as the posterior of evolutionary tree structure through Gibbs sampling; eventually outputs the ECMs of input gene set in the Partition step, and then classify novel genes into inferred ECMs in the Expansion step. 

Rather than using only a point tree estimate, CLIME 1.1 adds to the original CLIME 1.0 by allowing the estimation error in the tree-building process as well as the variability of phylogenetic trees among genes, and thus alleviating the risk of misspecification in the tree structure. In practice, popular tree-building methods and softwares such as PhyML \citep{guindon2010new} and MrBayes \citep{ronquist2003mrbayes} characterize the uncertainty in the estimation with  bootstrap or posterior tree samples. CLIME 1.1 can readily utilize such output samples as empirical approximation for tree prior distribution. We also compare CLIME 1.1 with CLIME 1.0 and other benchmark methods in extensive simulations and real data to showcase its features and strengths. We find that CLIME 1.1 is more robust and accurate when there is high uncertainty in tree estimation or gene-wise variability in the evolutionary tree structures. 

%CLIME uses a tree-structured hidden Markov model (HMM) to model the stochastic gain/loss process of genes on a given phylogenetic tree, and a Dirichlet process mixture (DPM) of HMMs is used for clustering input genes to co-evolved modules with shared evolutionary history. The posterior inference of the evolutionary history of genes and co-evolved gene module assignments is accomplished by dynamic programming and Markov chain Monte Carlo (MCMC). CLIME's HMM and DPM modeling gives it following advantages over the existing methods: (1) it can take into proper account of the dependency between species; (2) it automatically learns the number of distinct evolutionary modules in the input gene set $\mathcal{G}$; (3) it leverages information from the entire input gene set to more reliably predict new genes that have arisen with a shared pattern of evolutionary gains and losses; (4) it uses the LLR statistic of HMMs as much more statistical principled measure for co-evolution compared to naive metrics (e.g. Hamming distance, Pearson correlation).  

The rest of this article is organized as follows. In Section
\ref{sec:model}, we introduce the tree-structured hidden Markov model (HMM) for genes' stochastic gain/loss events on a given phylogenetic tree, and the Dirichlet process mixture (DPM) model for clustering genes
into modules with shared history. The Partition step of CLIME 1.0, which implements
the Gibbs sampler to sample from the posterior distribution of the
DPM model, is described in Section \ref{sec:partition}.
The Expansion step is introduced in Section \ref{sub:expansion}. In Section
\ref{sub:gain_null_est}, we briefly introduce the pre-processing
of CLIME 1.0. The extended model and inference procedure of CLIME 1.1 are described in Section \ref{sec:clime+}. Simulation studies that compare CLIME 1.0 and CLIME 1.1 with hierarchical
clustering  are presented in \ref{sec:simstudy}. In Section \ref{sec:realdata},
we apply CLIME 1.0 and 1.1 on real data, and use leave-one-out cross-validation
to compare the performance of CLIME 1.0 with hierarchical clustering on
gene sets from GO (Gene Ontology) and KEGG (Kyoto Encyclopedia of
Genes and Genomes) databases. We conclude this paper with a discussion
in Section \ref{sec:discussion}.

\section{Bayesian mixture of HMM on a phylogenetic tree}\label{sec:model}
\subsection{Notation}
Let $\mathcal{G}$ denote the input gene set
with $n$ genes, and $N$ be the total number of genes in the reference genome. Let $\mathbf{X}_i$ be the phylogenetic profile of gene $i$, $i=1, \dots, N$, and specifically, let $\mathbf{X}$ denote the phylogenetic profile of the input gene set. For example, $\mathcal{G}$
can be the set of 44 subunit genes of human mitochondrial complex I, and $\mathbf{X}$ is their phylogenetic profile matrix; for reference genome, we have $N=20,834$ human genes with their phylogenetic profile matrix denoted by $\mathbf{X}_{1:N}$.
For notational simplicity, we let $1,\dots,n$ index the $n$ genes
in $\mathcal{G}$ and let $n+1,\dots,N$ index the rest in the genome.
The input phylogenetic tree has $S$ living species
indexed by $1,\dots,S$, and $S-1$ ancestral extinct species indexed
by $S+1,\dots,2S-1$. The $2S-1$ living and extinct species are connected
by the $2S-2$ branches on the tree. For simplicity, we assume that
the phylogenetic tree is binary, while the model and algorithm can
be easily modified for non-binary input trees. For each gene $i=1,\dots,N$,
its {\it phylogenetic profile} is defined as the observed vector $\boldsymbol{X}_{i}=\left(X_{i,1},\dots,X_{i,S}\right)$ with
$X_{i,j}=1$ or $0$ denoting the presence or absence of gene $i$ 
across the $S$ extant species. Let $\boldsymbol{H}_{i}=\left(H_{i,1},\dots,H_{i,2S-1}\right)$ denote gene $i$'th ancestral (unobserved) and extant presence/absence states in the $2S-1$ species.

We call a cluster of genes with shared evolutionary history an
evolutionarily conserved module (ECM). Let $\boldsymbol{I}=\left(I_{1},\dots,I_{n}\right)$
denote the ECM assignment indicators of genes, where $I_{i}=k$ indicates
that gene $i$ is assigned to ECM $k$. We assume that each gene can
only be ``gained" once throughout the entire evolutionary history, which
happens at branch $\lambda_{i}$, $i=1,\dots,N$. Let $\boldsymbol{\lambda}=\left(\lambda_{1},\dots,\lambda_{N}\right)$
denote the gain nodes of the $N$ genes, where $\lambda_{i}=s$ indicates that
gene $i$ was gained at tree node $s$. With the available data, we can estimate $\boldsymbol{\lambda}$ in the pre-processing stage as described in Section \ref{sub:gain_null_est} with very small estimation error. We thus assume that $\boldsymbol{\lambda}$ is a known parameter throughout the main algorithm.
%and omit the symbol of conditioning on $\boldsymbol{\lambda}$ in likelihood functions. 

\subsection{Tree-structured HMM for phylogenetic profiles}\label{sub:ecm_mixture_model}

We introduce here a tree-structured HMM to model the presence/absence history
and phylogenetic profile of genes. For each gene $i$, its complete
evolutionary history $\boldsymbol{H}_{i}=\left(H_{i,1},\dots,H_{i,2S-1}\right)$ is only partially observed at the bottom level, i.e., the phylogenetic profile vector $\boldsymbol{X}_{i}=\left(X_{i,1},\dots,X_{i,S}\right)$
is the observation of presence/absence states for only the living species, 
$H_{i,1},\dots,H_{i,S}$. Due to sequencing and genome annotation
errors, there are also observation errors on the presence/absence
of genes. In other words, $X_{i,1},\dots,X_{i,S}$ are noisy observations
on $H_{i,1},\dots,H_{i,S}$. We assume that genes in ECM $k$ share
the same set of branch-specific probabilities of gene loss for the
$2S-2$ branches, denoted by $\boldsymbol{\theta}_{k}=\left(\theta_{k,1},\dots,\theta_{k,2S-2}\right)$.
For genes in ECM $k$, the transition of absence/presence states from
its direct ancestor to species $s$ is specified by transition matrix
$\mathbf{Q}_{k,s}$,\vspace{-10pt}
\[
\mathbf{Q}_{k,s}=\begin{array}{c}
\begin{array}{cc}
0 & \quad\;\;\:1\end{array}\\
\begin{array}{c}
0\\
1
\end{array}\left[\begin{array}{cc}
1 & 0\\
\theta_{k,s} & 1-\theta_{k,s}
\end{array}\right].
\end{array}
\]
Thus, for every evolutionary branch (after the gain branches $\mathbf{\lambda}$), there is a $\mathbf{Q}$ matrix. We assume that once a gene got lost, it cannot be re-gained, which
is realistic for eukaryotic species. Therefore the first row of $\mathbf{Q}_{k,s}$ indicates that the transition probability from absence to presence (re-gain) is $0$. The second
row shows our parameterization that the transition probability from
presence to absence (gene loss) is $\theta_{k,s}$, and presence to
presence is $1-\theta_{k,s}$. 

Let $\sigma\left(s\right)$ denote the direct ancestor species of
$s$, and let set $\mathcal{T}\left(s\right)$ include all of the
offspring species in the sub-tree rooted at node $s$. Obviously
$H_{i,s}=0$ if species $s$ is not in $\mathcal{T}\left(\lambda_{i}\right)$.
The likelihood function of evolutionary history $\boldsymbol{H}_{i}$
conditional on gene $i$ in ECM $k$ is 
\begin{eqnarray*}
 &  & \text{Pr}\left(\boldsymbol{H}_{i}\mid\boldsymbol{\theta}_{k},I_{i}=k\right)\\
 & = & \begin{cases}
\prod_{s\in\mathcal{T}\left(\lambda_{i}\right)\backslash\lambda_{i}}\mathbf{Q}_{k,s}\left(H_{i,\sigma\left(s\right)},H_{i,s}\right), & \text{if }\,H_{i,s}=0\,\forall s\not\in\mathcal{T}\left(\lambda_{i}\right),\\
0, & \text{otherwise}.
\end{cases}
\end{eqnarray*}

To account for errors in determining the presence/absence of a gene,
we allow each component of the observed phylogenetic profile, $X_{i,s}$,
to have an independent probability $q$ to be erroneous (i.e., different
from the true state $H_{i,s}$). The error probability $q$ is low and assumed to be known. By default, we set $q=0.01$
based on our communication with biologists with expertise in genome
sequencing and annotation. We note that estimating it in the MCMC
procedure is straightforward, but a strong prior on $q$ is
needed for its proper convergence and identifiability. For each gene $i$,
the likelihood function of $\boldsymbol{X}_{i}$ given $\boldsymbol{H}_{i}$
is
\begin{equation}
\text{Pr}\left(\boldsymbol{X}_{i}\mid\boldsymbol{H}_{i}\right)=\prod_{s=1}^{S}\text{Pr}\left(X_{i,s}\mid H_{i,s}\right)=\prod_{s=1}^{S}\left(1-q\right)^{\mathbb{I}\left\{ X_{i,s}=H_{i,s}\right\} }\left(q\right)^{\mathbb{I}\left\{ X_{i,s}\neq H_{i,s}\right\} },\label{eq:P(Xi|Hi)}
\end{equation}
where $\mathbb{I}\left\{ \cdot\right\} $ is the indicator function
that is equal to $1$ if the statement is true, and $0$ otherwise.
The complete likelihood for gene $i$ is
\begin{eqnarray}
 &  & \text{Pr}\left(\boldsymbol{X}_{i},\boldsymbol{H}_{i}\mid\boldsymbol{\theta},I_{i}\right)\nonumber  =  \label{eq:model_i} \\
 \!&  \! &\!\! \left[\prod_{s\in\mathcal{T}\left(\lambda_{i}\right)\backslash\lambda_{i}}\mathbf{Q}_{I_{i},s}\left(H_{i,\sigma\left(s\right)},H_{i,s}\right)\right]\left[\prod_{s=1}^{S}\left(1-q\right)^{\mathbb{I}\left\{ X_{i,s}=H_{i,s}\right\} }\left(q\right)^{\mathbb{I}\left\{ X_{i,s}\neq H_{i,s}\right\} }\right]
\end{eqnarray}
and the  complete likelihood for all the genes is
\begin{equation}
\text{Pr}\left(\boldsymbol{X},\boldsymbol{H}\mid\boldsymbol{\theta},\boldsymbol{I}\right)=\prod_{i=1}^{n}\text{Pr}\left(\boldsymbol{X}_{i},\boldsymbol{H}_{i}\mid\boldsymbol{\theta},I_{i}\right).
\end{equation}

\subsection{Dirichlet process mixture of tree hidden Markov models}

The number of ECMs $K$ may be specified by users reflecting their prior
knowledge on the data set. When the prior information about the data
set is not available, we can estimate $K$ from data by MCMC sampling
with a Dirichlet process prior on $\boldsymbol{\theta}$ \citep{ferguson1973abayesian,neal2000markovchain}.
For each gene $i\in\left\{ 1,\dots,n\right\} $, we let the prior
distribution of $\boldsymbol{\theta}_{i}$ follow Dirichlet process
with concentration parameter $\alpha$ and base distribution $\mathcal{F}_{0}$, denoted by $\text{DP}\left(\mathcal{F}_{0},\alpha\right)$.
This gives us the following Bayesian hierarchical model. For each
gene $i=1,\dots,n$,
\begin{equation}
\begin{aligned}\boldsymbol{X}_{i}\mid\boldsymbol{H}_{i} & \;\:\sim\;\:P\left(\boldsymbol{X}_{i}\mid\boldsymbol{H}_{i}\right),\\
\boldsymbol{H}_{i}\mid\boldsymbol{\theta}_{i} & \;\:\sim\;\:P\left(\boldsymbol{H}_{i}\mid\boldsymbol{\theta}_{i}\right),\\
\boldsymbol{\theta}_{i}\mid\mathcal{F} & \;\:\sim\;\:\mathcal{F},\\
\mathcal{F} & \;\:\sim\;\:\text{DP}\left(\mathcal{F}_{0},\alpha\right),\\
\mathcal{F}_{0} & \;\:=\;\:\prod_{s=1}^{2S-2}\text{Beta}\left(a,b\right),
\end{aligned}
\label{eq:DP}
\end{equation}
%where $\text{DP}\left(\mathcal{F}_{0},\alpha\right)$ is the Dirichlet
%process with base distribution $\mathcal{F}_{0}$ and scaling parameter
%$\alpha$. 
The base distribution $\mathcal{F}_{0}$ is set as the product
of a set of Beta distributions for branch-specific gene
loss probabilities. 

%The clustering based on Dirichlet process mixture model can be naturally
%implemented by MCMC sampling \citep{neal2000markovchain}. In particular,
We use the Chinese restaurant process representation \citep{aldous1985exchangeability,pitman1996somedevelopments}
of the Dirichlet process and implement a Gibbs sampler \citep{gelfand1990sampling,liu2008montecarlo}
to draw from the posterior distribution of ECM assignments $\boldsymbol{I}=\left(I_{1},\dots,I_{n}\right)$. The Chinese restaurant process prior for cluster assignments is exchangeable
\citep{aldous1985exchangeability}, therefore the prior distribution
for $\boldsymbol{I}$ is invariant to the order of $n$ genes.
More precisely, the
mixture model in Eq (\ref{eq:DP}) can be formulated as follows:
\begin{equation}
\begin{aligned}\boldsymbol{X}_{i}\mid\boldsymbol{H}_{i} & \;\:\sim\;\:P\left(\boldsymbol{X}_{i}\mid\boldsymbol{H}_{i}\right),\quad i=1,2,\dots n,\\
\boldsymbol{H}_{i}\mid\boldsymbol{\theta}_{I_{i}} & \;\:\sim\;\:P\left(\boldsymbol{H}_{i}\mid\boldsymbol{\theta}_{I_{i}}\right),\quad i=1,2,\dots n,\\
\boldsymbol{\theta}_{k} & \;\:\sim\;\:\prod_{s=1}^{2S-2}\text{Beta}\left(a,b\right),\quad k=1,2,\dots\\
\text{Pr}\left(I_{i}=I_{j},\;j<i\mid I_{1},\dots,I_{i-1}\right) & \;\:=\;\:n_{i,j}/\left(i-1+\alpha\right),\quad i=1,2,\dots n,\\
\text{Pr}\left(I_{i}\neq I_{j},\;\forall j<i\mid I_{1},\dots,I_{i-1}\right) & \;\:=\;\:\alpha/\left(i-1+\alpha\right),\quad i=1,2,\dots n,
\end{aligned}
\label{eq:CRP}
\end{equation}
where $n_{i,j}=\sum_{l=1}^{i-1}\mathbb{I}\left\{ I_{l}=I_{j}\right\} $.

\subsection{Dynamic programming for integrating out $\boldsymbol{H}$\label{sub:H_integration}}

In Section \ref{sub:gibbs}, we will introduce the Gibbs sampler to
sample from the posterior distribution of $\boldsymbol{I}$. In the
Gibbs sampler, we need to calculate the marginal probability of $\mathbf{X}_i$ given the HMM parameter $\boldsymbol{\theta}$, with  gene $i$'s  
 evolutionary history $\boldsymbol{H}_{i}$ integrated out. Suppose gene $i$ is in ECM $k$, then
\[
\text{Pr}\left(\boldsymbol{X}_{i}\mid\boldsymbol{\theta}_{k}\right)\,\,=\,\,\sum_{\boldsymbol{H}_{i}}\text{Pr}\left(\boldsymbol{X}_{i},\boldsymbol{H}_{i}\mid\boldsymbol{\theta}_{k}\right).
\]
We use the following tree-version of the backward procedure to calculate
this marginal probability. For gene $i$, define
$\boldsymbol{X}_{i}^{s}$ as its phylogenetic profile in the sub-tree
rooted at species $s$ (obviously $\boldsymbol{X}_{i}^{2S-1}=\boldsymbol{X}_{i}$).
We calculate the marginal probability by recursively computing factors
$\beta_{i,s}\left(h\right)$, defined as 
\[
\beta_{i,s}\left(h\right)\,\,\equiv\,\,\text{Pr}\left(\boldsymbol{X}_{i}^{s}\mid\boldsymbol{\theta}_{k},H_{i,s}=h\right).
\]
For a living species $s$, which is a leaf of the tree, 
\[
\beta_{i,s}\left(h\right)\,\,=\,\,\text{Pr}\left(X_{i}^{s}\mid\boldsymbol{\theta}_{k},H_{i,s}=h\right)\,\,=\,\,\left(1-q\right)^{\mathbb{I}\left\{ X_{i}^{s}=h\right\} }\left(q\right)^{\mathbb{I}\left\{ X_{i}^{s}\neq h\right\} }.
\]
Let $\delta_{1}\left(s\right)$ and $\delta_{2}\left(s\right)$ denote
those two children species of $s$. For a inner tree species $s$,
we can factorize $\beta_{i,s}\left(t\right)$ as
\begin{eqnarray*}
\beta_{i,s}\left(h\right)
 & = & \sum_{h_{1},h_{2}\in\left\{ 0,1\right\} }\text{Pr}\left(\boldsymbol{X}_{i}^{s},H_{i,\delta_{1}\left(s\right)}=h_{1},H_{i,\delta_{2}\left(s\right)}=h_{2}\mid\boldsymbol{\theta}_{k},H_{i,s}=h\right)\\
 & = & \sum_{h_{1},h_{2}\in\left\{ 0,1\right\} }\text{Pr}\left(\boldsymbol{X}_{i}^{s}\mid\boldsymbol{\theta}_{k},H_{i,\delta_{1}\left(s\right)}=h_{1}\right)\cdot\text{Pr}\left(H_{i,\delta_{1}\left(s\right)}=h_{1}\mid\boldsymbol{\theta}_{k},H_{i,s}=h\right)\\
 &  & \cdot\text{Pr}\left(\boldsymbol{X}_{i}^{s}\mid\boldsymbol{\theta}_{k},H_{i,\delta_{2}\left(s\right)}=h_{2}\right)\cdot\text{Pr}\left(H_{i,\delta_{1}\left(s\right)}=h_{2}\mid\boldsymbol{\theta}_{k},H_{i,s}=h\right)\\
 & = & \left[\sum_{h_{1}\in\left\{ 0,1\right\} }\beta_{i,\delta_{1}\left(s\right)}\left(h_{1}\right)\mathbf{Q}_{k,\delta_{1}\left(s\right)}\left(h,h_{1}\right)\right]\left[\sum_{h_{2}\in\left\{ 0,1\right\} }\beta_{i,\delta_{2}\left(s\right)}\left(h_{2}\right)\mathbf{Q}_{k,\delta_{2}\left(s\right)}\left(h,h_{2}\right)\right].
\end{eqnarray*}
For each gene $i$, we calculate the $\beta$'s recursively bottom-up along
the tree, until the gain branch $\lambda_{i}$, resulting in  the marginal
probability: 
%$\text{Pr}\left(\boldsymbol{X}_{i}\mid\boldsymbol{\theta}_{k}\right)$ can be written as
%is $\beta_{i,\lambda_{i}}\left(1\right)$ as shown below. 
\begin{eqnarray}
\text{Pr}\left(\boldsymbol{X}_{i}\mid\boldsymbol{\theta}_{k}\right) & = & \sum_{h\in\left\{ 0,1\right\} }\text{Pr}\left(\boldsymbol{X}_{i}^{\lambda_{i}}\mid\boldsymbol{\theta}_{k},H_{i,\lambda_{i}}=h\right)\text{Pr}\left(H_{i,\lambda_{i}}=h\mid\boldsymbol{\theta}_{k}\right)\nonumber \\
 & = & 0+\text{Pr}\left(\boldsymbol{X}_{i}^{\lambda_{i}}\mid\boldsymbol{\theta}_{k},H_{i,\lambda_{i}}=1\right)\:\stackrel{def}{=} \:\beta_{i,\lambda_{i}}\left(1\right).\label{eq:P(X|theta)}
\end{eqnarray}

\subsection{Dynamic programming for integrating out $\boldsymbol{\theta}$\label{sub:theta_integration}}

In each step of the Gibbs sampler, we pull out each gene from its
current ECM and either re-assign it to an existing ECM or create a new singleton ECM for it according to the calculated conditional probability $\text{Pr}\left(I_{i}\mid\boldsymbol{X}_{i},\boldsymbol{H}_{i},\boldsymbol{\theta}\right)$.
For each ECM $k$, its parameter $\boldsymbol{\theta}_{k}=\{\theta_{k,s}\}_{s=1}^{2S-2}$ is a vector containing $2S-2$ loss probabilities.
Our real data has $S=139$, which makes each $\boldsymbol{\theta}_{k}$
a $276$-dimensional vector. The high dimensionality of 
$\boldsymbol{\theta}_{1},\dots,\boldsymbol{\theta}_{K}$ adds heavy
computational burden and dramatically slows down the convergence rate
of the Gibbs sampler. To overcome this difficulty, we develop a collapsed
Gibbs sampler \citep{liu1994collapsed} by applying the predictive
updating technique \citep{chen1996predictive} to improve the MCMC
sampling efficiency. In particular, we integrate $\boldsymbol{\theta}_{k}$
out from the conditional probability $\text{Pr}\left(I_{i}=k\mid\boldsymbol{X}_{i},\boldsymbol{H}_{i},\boldsymbol{\theta}_{k}\right)$, so that
\begin{eqnarray*}
\text{Pr}\left(I_{i}=k\mid\boldsymbol{X}_{i},\boldsymbol{H},\boldsymbol{I}_{-i}\right) & = & \int\text{Pr}\left(I_{i}=k\mid\boldsymbol{X}_{i},\boldsymbol{H},\boldsymbol{\theta}_{k}\right)\text{Pr}\left(\boldsymbol{\theta}_{k}\mid\boldsymbol{X}_{i},\boldsymbol{H},\boldsymbol{I}_{-i}\right)d\boldsymbol{\theta}_{k}\\
 & \propto & \text{Pr}\left(\boldsymbol{X}_{i}\mid\boldsymbol{H}_{-i}^{k},I_{i}=k\right)\text{Pr}\left(I_{i}=k\mid\boldsymbol{I}_{-i}\right),
\end{eqnarray*}
where $\boldsymbol{H}^{k}=\left\{ \boldsymbol{H}_{j}:\,I_{j}=k,\,j=1,\dots,n\right\} $
denotes the evolutionary histories for genes in ECM $k$, and $\boldsymbol{H}_{-i}^{k}=\boldsymbol{H}^{k}\backslash\left\{ \boldsymbol{H}_{i}\right\} $. $\text{Pr}\left(I_{i}=k\mid\boldsymbol{I}_{-i}\right)\\ =\sum_{j\neq i}\mathbb{I}\left\{ I_{j}=k\right\} /\left(n-1+\alpha\right)$
is the Chinese restaurant prior on $\boldsymbol{I}$, and $\text{Pr}\left(\boldsymbol{X}_{i}\mid\boldsymbol{H}_{-i}^{k},I_{i}=k\right)$
is the marginal likelihood of $\boldsymbol{X}_{i}$ conditional on
gene $i$ is in ECM $k$ with $\boldsymbol{\theta}_{k}$ integrated
out. We calculate $\text{Pr}\left(\boldsymbol{X}_{i}\mid\boldsymbol{H}_{-i}^{k},I_{i}=k\right)$
as follows.

Conditional on $\boldsymbol{H}_{-i}^{k}$, the distribution of $\theta_{k,s}$,
$s=1,\dots,2S-2$, is simply a conjugate Beta posterior distribution,
\begin{eqnarray*}
\theta_{k,s}\mid\boldsymbol{H}_{-i}^{k} & \sim & \text{Beta}\left(a+\sum_{j\neq i,I_{j}=k}\mathbb{I}\left\{ H_{j,\sigma\left(s\right)}=1,H_{j,s}=0\right\} ,\right.\\
 &  & \left.b+\sum_{j\neq i,I_{j}=k}\mathbb{I}\left\{ H_{j,\sigma\left(s\right)}=1,H_{j,s}=1\right\} \right).
\end{eqnarray*}
Integrating out $\boldsymbol{\theta}_{k}$ with respect to this distribution,
we obtain the likelihood of $\boldsymbol{X}_{i}$ conditional on $\boldsymbol{H}_{-i}^{k}$:
\begin{eqnarray}
\text{Pr}\left(\boldsymbol{X}_{i}\mid\boldsymbol{H}_{-i}^{k},I_{i}=k\right) & = & \int\text{Pr}\left(\boldsymbol{X}_{i}\mid\boldsymbol{\theta}_{k},I_{i}=k\right)\text{Pr}\left(\boldsymbol{\theta}_{k}\mid\boldsymbol{H}_{-i}^{k}\right)d\boldsymbol{\theta}_{k}\nonumber \\
 & = & \int\beta_{i,\lambda_{i}}\left(1\right)\text{Pr}\left(\boldsymbol{\theta}_{k}\mid\boldsymbol{H}_{-i}^{k}\right)d\boldsymbol{\theta}_{k}\:=\:\bar{\beta}_{i,\lambda_{i}}\left(1\right),\label{eq:P(Xi|H,Ii=00003Dk)}
\end{eqnarray}
where $\bar{\beta}$ is defined as 
\[
\bar{\beta}_{i,s}\left(h\right)\equiv\mathbb{E}\left[\beta_{i,s}\left(h\right)\mid\boldsymbol{H}_{-i}^{k}\right]=\mathbb{E}\left[\text{Pr}\left(\boldsymbol{X}_{i}^{s}\mid\boldsymbol{\theta}_{k},H_{i,s}=h\right)\mid\boldsymbol{H}_{-i}^{k}\right].
\]
For a leaf species $s$, $\bar{\beta}_{i,s}\left(h\right)=\beta_{i,s}\left(h\right)$.
For an inner tree species $s$, $\bar{\beta}_{i,s}\left(h\right)$
can be calculated recursively from bottom of the tree to the top as
\begin{eqnarray*}
 &  & \bar{\beta}_{i,s}\left(h\right)
  =  \mathbb{E}\left[\beta_{i,s}\left(h\right)\mid\boldsymbol{H}_{-i}^{k}\right]\\
 & = & \left[\sum_{h_{1}=0,1 }\bar{\beta}_{i,\delta_{1}\left(s\right)}\left(h_{1}\right)\bar{\mathbf{Q}}_{k,\delta_{1}\left(s\right)}\left(h,h_{1}\right)\right]\left[\sum_{h_{2}=0,1}\bar{\beta}_{i,\delta_{2}\left(s\right)}\left(h_{2}\right)\bar{\mathbf{Q}}_{k,\delta_{2}\left(s\right)}\left(h,h_{2}\right)\right].
\end{eqnarray*}
where $\bar{\mathbf{Q}}_{k,s}$ is the expectation of transition probability
matrix $\mathbf{Q}_{k,s}$ conditional on $\boldsymbol{H}_{-i}^{k}$,
\begin{eqnarray}
\bar{\mathbf{Q}}_{k,s} & = & \mathbb{E}\left[\mathbf{Q}_{k,s}\mid\boldsymbol{H}_{-i}^{k}\right]=\left[\begin{array}{cc}
1 & 0\\
\mathbb{E}\left[\theta_{k,s}\mid\boldsymbol{H}_{-i}^{k}\right] & 1-\mathbb{E}\left[\theta_{k,s}\mid\boldsymbol{H}_{-i}^{k}\right]
\end{array}\right],\label{eq:Q_bar}
\end{eqnarray}
and $\mathbb{E}\left[\theta_{k,s}\mid\boldsymbol{H}_{-i}^{k}\right]$
is simply the expectation of a Beta conjugate posterior distribution.
\[
\mathbb{E}\left[\theta_{k,s}\mid\boldsymbol{H}_{-i}^{k}\right]=\frac{a+\sum_{j:\,I_{j}=k,\,j\neq i}\mathbb{I}\left\{ H_{j,\delta\left(s\right)}=1,H_{j,s}=0\right\} }{a+b+\sum_{j:\,I_{j}=k,\,j\neq i}\mathbb{I}\left\{ H_{j,\delta\left(s\right)}=1\right\} }.
\]

In the Gibbs sampler, we also need to compute the marginal probability
that gene $i$ is in its own singleton group, i.e. $\text{Pr}\left(\boldsymbol{X}_{i}\mid I_{i}\neq I_{j},\,\forall j\neq i\right)$.
By integrating out $\boldsymbol{H}_{i}$ and $\boldsymbol{\theta}_{i}$,
we have 
\begin{eqnarray}
\text{Pr}\left(\boldsymbol{X}_{i}\mid I_{i}\neq I_{j},\,\forall j\neq i\right) & = & \int\sum_{\boldsymbol{H}_{i}}\text{Pr}\left(\boldsymbol{X}_{i},\boldsymbol{\theta}_{i},\boldsymbol{H}_{i}\mid I_{i}\neq I_{j},\,\forall j\neq i\right)d\boldsymbol{\theta}_{i}\nonumber \\
 & = & \int\text{Pr}\left(\boldsymbol{X}_{i}\mid\boldsymbol{\theta}_{i},I_{i}\neq I_{j},\,\forall j\neq i\right)d\mathcal{F}_{0}\left(\boldsymbol{\theta}_{i}\right)\nonumber \\
 & = & \int\beta_{i,\lambda_{i}}\left(1\right)\text{Pr}\left(\boldsymbol{\theta}_{i}\right)d\boldsymbol{\theta}_{i}.\label{eq:Pr(Xi|Ii=00003DK+1)}
\end{eqnarray}
Note that (\ref{eq:Pr(Xi|Ii=00003DK+1)}) is a special case of (\ref{eq:P(Xi|H,Ii=00003Dk)})
with $\boldsymbol{H}_{-i}^{k}=\emptyset$, thus it can be calculated
in the same recursive way with
\begin{eqnarray*}
\bar{\mathbf{Q}}_{k,s} & = & \mathbb{E}\left[\mathbf{Q}_{k,s}\mid\boldsymbol{H}_{-i}^{k}=\emptyset\right]=\left[\begin{array}{cc}
1 & 0\\
a/\left(a+b\right) & b/\left(a+b\right)
\end{array}\right].
\end{eqnarray*}

\subsection{ECM strength measurement}

After partitioning the input gene set $\mathcal{G}$ into ECMs, it
is of great interest to determine which of the ECMs share more informative
and coherent evolutionary histories than others, since the ranking
of ECMs leads to different priorities for further low-throughput experimental
investigations. In our Bayesian model-based framework, the strength
of ECM $k$, denoted by $\phi_{k}$, is defined as the logarithm of
the Bayes Factor between two models normalized by the number of genes
in that ECM. The first model is under the assumption that these genes
have co-evolved in the same ECM and share the same $\boldsymbol{\theta}$
parameter, and the second model is under the assumption that each
gene has evolved independently in its own singleton ECM with different
$\boldsymbol{\theta}$s. Specifically, with a partitioning configuration
$\boldsymbol{I}$, the strength for ECM $k$ is defined as 
\begin{equation}
\phi_{k}\,=\,\left\{ \log\left[\frac{\int\left[\prod_{i:\,I_{i}=k}\text{Pr}\left(\boldsymbol{X}_{i}\mid\boldsymbol{\theta}\right)\right]\text{Pr}\left(\boldsymbol{\theta}\right)d\boldsymbol{\theta}}{\prod_{i:\,I_{i}=k}\int\text{Pr}\left(\boldsymbol{X}_{i}\mid\boldsymbol{\theta}\right)\text{Pr}\left(\boldsymbol{\theta}\right)d\boldsymbol{\theta}}\right]\right\} \,/\:\sum_{i=1}^{n}\mathbb{I}\left\{ I_{i}=k\right\} .
\end{equation}
This strength measurement reflects the level of homogeneity among
the evolutionary histories of genes in this ECM. A larger $\phi_{k}$
indicates that genes in ECM $k$ share more similar and informative
evolutionary history with more branches having high loss probabilities.

\section{Partition step: MCMC sampling and point estimators\label{sec:partition}}

\subsection{Choice of hyper-parameters}

Several hyper-parameters need to be specified, including the concentration
parameter $\alpha$ in the Dirichlet process prior and hyper-parameters
$a,b$ for the Beta prior of $\theta$s. Concentration parameter $\alpha$
controls the prior belief for the number of components in the mixture
model, as larger $\alpha$ makes it easier to create a new ECM in
each step of the Gibbs sampling. We set Dirichlet process concentration
parameter as widely used $\alpha=1$. To test the method's robustness
on $\alpha$, we applied the algorithm to simulated and real data
with $\alpha=1$, $\alpha=\log\left(n\right)$ and $\alpha=\sqrt{n}$
respectively, and observed no significant changes on the posterior
distribution of $K$. The reason is that histories of ECMs are often
so different from each other that the likelihood function dominates
the prior on determining $K$.

We set hyper-parameters $\alpha=0.03$, $\beta=0.97$ to make the
prior have mean $0.03$, which reflects our belief that overall $3\%$
of times a gene gets lost when evolving from one species to another
on a branch of the tree. The $3\%$ average loss probability was determined
based on the genome-wide average loss rate observed in our data.

\subsection{Forwad-backward sampling for $\boldsymbol{H}$\label{sub:H_sampling}}

In the Gibbs sampler, we apply a tree-version of forward-summation-backward-sampling
method \citep[Sec. 2.4]{liu2008montecarlo} to sample/impute the hidden
evolutionary history states in $\boldsymbol{H}$. Conditional on gene
$i$ is in ECM $k$, we want to sample $\boldsymbol{H}_{i}$ from
the conditional distribution $\text{Pr}\left(\boldsymbol{H}_{i}\mid\boldsymbol{X}_{i},\boldsymbol{\theta}_{k}\right)$.
Note that, by the Markovian structure of tree HMM, $\text{Pr}\left(\boldsymbol{H}_{i}\mid\boldsymbol{X}_{i},\boldsymbol{\theta}_{k}\right)$
can be written as 
\begin{eqnarray*}
 &  & \text{Pr}\left(\boldsymbol{H}_{i}\mid\boldsymbol{X}_{i},\boldsymbol{\theta}_{k}\right)\\
 & = & \begin{cases}
\prod_{s\in\mathcal{T}\left(\lambda_{i}\right)\backslash\lambda_{i}}\text{Pr}\left(H_{i,s}\mid H_{i,\sigma\left(s\right)},\boldsymbol{X}_{i},\boldsymbol{\theta}_{k}\right) & \text{if }H_{i,s}=0\,\,\forall s\not\in\mathcal{T}\left(\lambda_{i}\right),\\
0 & \text{otherwise}.
\end{cases}
\end{eqnarray*}
which suggests a sequential sampling procedure: draw $H_{i,s}$ for
each species $s\in\mathcal{T}\left(\lambda_{i}\right)\backslash\lambda_{i}$
top-down along the tree from $\text{Pr}\left(H_{i,s}\mid H_{i,\sigma\left(s\right)},\boldsymbol{X}_{i},\boldsymbol{\theta}_{k}\right)$
conditional on the previously drawn state $H_{i,\sigma\left(s\right)}$
of its ancestral species $\sigma\left(s\right)$.

We first use the backward procedure described in Section (\ref{sub:H_integration})
to calculate the $\beta_{i,s}$ for all species $s\in\mathcal{T}\left(\lambda_{i}\right)\backslash\lambda_{i}$
bottom-up along the tree, then we have
%$\text{Pr}\left(H_{i,s}\mid H_{i,\sigma\left(s\right)},\boldsymbol{X}_{i},\boldsymbol{\theta}_{k}\right)$ can be calculated as 
\begin{eqnarray*}
\text{Pr}\left(H_{i,s}\mid H_{i,\sigma\left(s\right)},\boldsymbol{X}_{i},\boldsymbol{\theta}_{k}\right) & \propto & \text{Pr}\left(H_{i,s},\boldsymbol{X}_{i}^{s}\mid H_{i,\sigma\left(s\right)},\boldsymbol{\theta}_{k}\right)\\
 & = & \text{Pr}\left(\boldsymbol{X}_{i}^{s}\mid H_{i,s},\boldsymbol{\theta}_{k}\right)\cdot\text{Pr}\left(H_{i,s}\mid H_{i,\sigma\left(s\right)},\boldsymbol{\theta}_{k}\right)\\
 & = & \beta_{i,s}\left(H_{i,s}\right)\cdot\mathbf{Q}_{k,s}\left(H_{i,\sigma\left(s\right)},H_{i,s}\right).
\end{eqnarray*}
Similar to Section \ref{sub:theta_integration}, we integrate out
$\boldsymbol{\theta}_{k}$ to derive that 
%from the complete likelihood $\text{Pr}\left(\boldsymbol{X}_{i},\boldsymbol{H}_{i}\mid\boldsymbol{\theta}_{k}\right)$ over $p\left(\boldsymbol{\theta}_{k}\mid\boldsymbol{H}_{-i},I_{i}=k\right)$ as follows,
\begin{eqnarray}
 &  & \text{Pr}\left(\boldsymbol{X}_{i},\boldsymbol{H}_{i}\mid\boldsymbol{H}_{-i},I_{i}=k\right) \label{eq:P(X,H|H-i,I)} \\
 &  {=} &  \int\text{Pr}\left(\boldsymbol{X}_{i},\boldsymbol{H}_{i}\mid\boldsymbol{\theta}_{k}\right)\text{Pr}\left(\boldsymbol{\theta}_{k}\mid\boldsymbol{H}_{-i},I_{i}=k\right)d\boldsymbol{\theta}_{k} \nonumber \\
 & {=} &  \left[\prod_{s\in\mathcal{T}\left(\lambda_{i}\right)\backslash\lambda_{i}} \!\bar{\mathbf{Q}}_{k,s}\left(H_{i,\sigma\left(s\right)},H_{i,s}\right)\right]\left[\prod_{s=1}^{S}\left(1-q\right)^{\mathbb{I}\left\{ X_{i,s}=H_{i,s}\right\} }q^{\mathbb{I}\left\{ X_{i,s}\neq H_{i,s}\right\} }\right] \nonumber,
\end{eqnarray}
where $\bar{\mathbf{Q}}_{k,s}$ was defined in Eq (\ref{eq:Q_bar}).
Obviously, Eq (\ref{eq:P(X,H|H-i,I)}) is in the same form as the complete
likelihood in Eq (\ref{eq:model_i}) with transition probabilities matrix
$\mathbf{Q}_{k,s}$ replaced by $\bar{\mathbf{Q}}_{k,s}$. The sequential
sampling strategy for $\boldsymbol{H}_{i}$ from conditional distribution
$\text{Pr}\left(\boldsymbol{H}_{i}\mid\boldsymbol{X}_{i},\boldsymbol{H}_{-i},I_{i}=k\right)$
is to start with $H_{i,\lambda_{i}}=1$ and draw $H_{i,s}$ for each
species $s\in\mathcal{T}\left(\lambda_{i}\right)\backslash\lambda_{i}$
top-down along the tree from distribution $\text{Pr}\left(H_{i,s}\mid H_{i,\sigma\left(s\right)},\boldsymbol{X}_{i},\boldsymbol{H}_{-i},I_{i}=k\right)$
conditional on the sampled state $H_{i,\sigma\left(s\right)}$ of
its ancestral species $\sigma\left(s\right)$, with matrices $\mathbf{Q}_{k,s}$
replaced by $\bar{\mathbf{Q}}_{k,s}$.

\subsection{Gibbs sampling implementation\label{sub:gibbs}}

In each step of Gibbs sampling, we pull out each gene from its current
ECM and assign it to an existing ECM or create a new singleton ECM
for it with respect to the calculated conditional distribution $\text{Pr}\left(I_{i}\mid\boldsymbol{X}_{i},\boldsymbol{H},\boldsymbol{I}_{-i}\right)$,
which is calculated as 
\begin{eqnarray}
 &  & \text{Pr}\left(I_{i}=k\mid\boldsymbol{X}_{i},\boldsymbol{H},\boldsymbol{I}_{-i}\right)\label{eq:P(I=00003Dk|X,H,I)}\\
 & \propto & \begin{cases}
\frac{\sum_{j:\,j\neq i}\mathbb{I}\left\{ I_{j}=k\right\} }{n-1+\alpha}\cdot\text{Pr}\left(\boldsymbol{X}_{i}\mid\boldsymbol{H}_{-i},I_{i}=k\right), & \exists j\neq i,\text{ s.t. }I_{j}=k,\\
\frac{\alpha}{n-1+\alpha}\cdot\text{Pr}\left(\boldsymbol{X}_{i}\mid I_{i}\neq I_{j},\,\forall j\neq i\right), & \text{otherwise}.
\end{cases}\nonumber 
\end{eqnarray}
where $\text{Pr}\left(\boldsymbol{X}_{i}\mid\boldsymbol{H}_{-i},I_{i}=k\right)$
and $\text{Pr}\left(\boldsymbol{X}_{i}\mid I_{i}\neq I_{j},\,\forall j\neq i\right)$
are respectively calculated in Eqs (\ref{eq:P(Xi|H,Ii=00003Dk)})
and (\ref{eq:Pr(Xi|Ii=00003DK+1)}).

We implement the collapsed Gibbs sampler to calculate the posterior
distribution of $\boldsymbol{I}$ and $\boldsymbol{H}$. In each Gibbs
sampler iteration, we conduct the following two steps: 
\begin{enumerate}
\item Draw $\boldsymbol{H}_{i}\sim\text{Pr}\left(\boldsymbol{H}_{i}\mid\boldsymbol{X}_{i},\boldsymbol{H}_{-i},\boldsymbol{I}\right),\;i=1,\dots,n$
by the procedure in Section \ref{sub:H_sampling}.
\item Draw $I_{i}\sim\text{Pr}\left(I_{i}\mid\boldsymbol{X}_{i},\boldsymbol{H},\boldsymbol{I}_{-i}\right),\;i=1,\dots,n$
as calculated in Eq (\ref{eq:P(I=00003Dk|X,H,I)}).
\end{enumerate}
By using this Gibbs sampling scheme, genes with similar evolutionary
history will be clustered to the same ECM, and genes without any close
neighbor will stay in their own singleton ECMs. This automatically
estimates the number of ECMs $K$. 

We implemented this Gibbs sampler in C++, and tested its computational
efficiency. On a typical input gene set with $\sim$ $100$
genes across $139$ species, the Gibbs sampler takes about 30
minutes to finish $1000$ iterations on a standard Linux server using
a single CPU. For input gene sets  of size $5000$, the Gibbs sampler
takes less than 24 hours to finish $1000$ iterations.

\subsection{Point estimator for ECM assignments $\boldsymbol{I}$\label{sub:ml_calc}}

While the posterior distribution of $\boldsymbol{I}$ is calculated
by the Gibbs sampler, users may prefer a single optimal solution for
$\boldsymbol{I}$ as it is easier to interpret and proceed to further
experimental investigations. To obtain a point estimator
of $\boldsymbol{I}$, we calculate the posterior probability $\text{Pr}\left(\boldsymbol{I}\mid\boldsymbol{X}\right)$
at the end of each Gibbs sampling iteration. The {\it maximum a posteriori} (MAP) assignment,
%and ECM assignments $\boldsymbol{I}$
%with highest 
$\arg\max_{\boldsymbol{I}}\text{Pr}\left(\boldsymbol{I}\mid\boldsymbol{X}\right)$,
%denoted as $\hat{\boldsymbol{I}}$,  
will be reported as the final MAP estimation. Suppose we have $M$ MCMC samples, denoted
by $\boldsymbol{I}^{\left(1\right)},\dots,\boldsymbol{I}^{\left(M\right)}$,
then the MAP assignment can be approximated by
\[
\hat{\boldsymbol{I}}\,=\,\underset{\boldsymbol{I}^{\left(m\right)}:\,m=1,\dots,M}{\arg\max}\:\text{Pr}\left(\boldsymbol{I}^{\left(m\right)}\mid\boldsymbol{X}\right).
\]
We know that 
\[
\text{Pr}\left(\boldsymbol{I}\mid\boldsymbol{X}\right)\,\propto\,\text{Pr}\left(\boldsymbol{X}\mid\boldsymbol{I}\right)\text{Pr}\left(\boldsymbol{I}\right),
\]
where $\text{Pr}\left(\boldsymbol{I}\right)$ is the Chinese restaurant
process prior, 
\[
\text{Pr}\left(\boldsymbol{I}\right)=\frac{\prod_{k=1}^{K}\left(n_{k}-1\right)!}{n!},\quad\,\text{where }\,\,n_{k}=\sum_{i=1}^{n}\mathbb{I}\left\{ I_{i}=k\right\} ,
\]
and $\text{Pr}\left(\boldsymbol{X}\mid\boldsymbol{I}\right)=\prod_{k=1}^{K}\text{Pr}\left(\boldsymbol{X}_{k}\mid\boldsymbol{I}\right)$,
where $\boldsymbol{X}_{k}=\left\{ \boldsymbol{X}_{i}:\,I_{i}=k,i=1,\dots,n\right\} $ and $\text{Pr}\left(\boldsymbol{X}_{k}\mid\boldsymbol{I}\right)$
is the marginal probability for phylogenetic profiles of genes in
ECM $k$, i.e.,
\begin{eqnarray*}
\text{Pr}\left(\boldsymbol{X}_{k}\mid\boldsymbol{I}\right) & = & \int\left[\prod_{i:I_{i}=k}\text{Pr}\left(\boldsymbol{X}_{i}\mid\boldsymbol{\theta}_{k}\right)\right]\text{Pr}\left(\boldsymbol{\theta}_{k}\right)d\boldsymbol{\theta}_{k}.
\end{eqnarray*}
%However, this integral has no closed-form solution and we use Monte Carlo samples drawn from Gibbs sampling to approximate
This integral has no closed-form solution, but we can approximate this marginal likelihood by the method in \citet{chib1995marginal} using samples obtained by the Gibbs sampler.
In particular, we have the following equation holds for any $\boldsymbol{\theta}_{k}^{*}=\left(\theta_{k,1}^{*},\dots,\theta_{k,2S-1}^{*}\right)$:
\begin{equation}
\log\text{Pr}\left(\boldsymbol{X}_{k}\mid\boldsymbol{I}\right) \! = \! \sum_{i:I_{i}=k}\log \text{Pr}\left(\boldsymbol{X}_{i}\mid\boldsymbol{\theta}_{k}^{*}\right)+\log \text{Pr}\left(\boldsymbol{\theta}_{k}^{*}\right)-\log \text{Pr}\left(\boldsymbol{\theta}_{k}^{*}\mid\boldsymbol{X}_{k},\boldsymbol{I}\right).
\label{eq:lnp(X_k|I)}
\end{equation}
In the equation above, prior probability $\text{Pr}\left(\boldsymbol{\theta}_{k}^{*}\right)$
can be calculated directly and the likelihood $\text{Pr}\left(\boldsymbol{X}_{i}\mid\boldsymbol{\theta}_{k}^{*}\right)$
can be calculated by dynamic programming with computational complexity
$O\left(S\right)$. We approximate $\text{Pr}\left(\boldsymbol{\theta}_{k}^{*}\mid\boldsymbol{X}_{k},\boldsymbol{I}\right)$
by running additional Gibbs sampling. Let $\boldsymbol{H}_{k}=\left\{ \boldsymbol{H}_{i}:\,I_{i}=k,\,i=1,\dots,n\right\} $.
We fix ECM assignments at $\boldsymbol{I}$ and re-run Gibbs sampler
for $T$ iterations to draw samples $\left\{ \boldsymbol{H}_{k}^{\left(1\right)},\dots,\boldsymbol{H}_{k}^{\left(M\right)}\right\} $
from $\text{Pr}\left(\boldsymbol{H}_{k}\mid\boldsymbol{X}_{k},\boldsymbol{I}\right)$,
and then $\text{Pr}\left(\boldsymbol{\theta}_{k}^{*}\mid\boldsymbol{X}_{k},\boldsymbol{I}\right)$
can be approximated as
\begin{equation}
\text{Pr}\left(\boldsymbol{\theta}_{k}^{*}|\boldsymbol{X}_{k},\boldsymbol{I}\right)=\sum_{\boldsymbol{H}_{k}}\text{Pr}\left(\boldsymbol{\theta}_{k}^{*}|\boldsymbol{H}_{k}\right)\text{Pr}\left(\boldsymbol{H}_{k}|\boldsymbol{X}_{k},\boldsymbol{I}\right)\approx\frac{1}{M}\sum_{m=1}^{M}\text{Pr}\left(\boldsymbol{\theta}_{k}^{*} | \boldsymbol{H}_{k}^{\left(m\right)}\right),\label{eq:p(theta_k|X_k, I)}
\end{equation}
where
\begin{eqnarray*}
\text{Pr}\left(\boldsymbol{\theta}_{k}^{*}\mid\boldsymbol{H}_{k}^{\left(m\right)}\right)
 &{=}& \prod_{s=1}^{2S-2}\text{Be}\left(\theta_{k,s}^{*}\left\vert a+\sum_{i:\,I_{i}=k}\mathbb{I}\left\{ H_{i,\delta\left(s\right)}^{\left(m\right)}=1,H_{i,s}^{\left(m\right)}=0\right\} ,\right.\right.\\
 &  & \quad\quad\quad\quad\quad\left. b+\sum_{i:\,I_{i}=k}\mathbb{I}\left\{ H_{i,\delta\left(s\right)}^{\left(m\right)}=1,H_{i,s}^{\left(m\right)}=1\right\} \right).
\end{eqnarray*}
$\text{Be}(\theta | \alpha,\beta)$
%={\Gamma(\alpha+\beta) \over \Gamma(\alpha) \ \Gamma(\beta)} \theta^{\alpha-1}(1-\theta)^{\beta-1}$ 
is the Beta density function.
%\begin{eqnarray*}
%\theta_{k,s}^{*}|\boldsymbol{H}_{k}^{\left(t\right)}\sim Beta( a+\sum_{i:\,I_{i}=k}\mathbb{I}\left\{ H_{i,\delta\left(s\right)}^{\left(t\right)}=1,H_{i,s}^{\left(t\right)}=0\right\} ,\\
% b+\sum_{i:\,I_{i}=k}\mathbb{I}\left\{ H_{i,\delta\left(s\right)}^{\left(t\right)}=1,H_{i,s}^{\left(t\right)}=1\right\}).
%\end{eqnarray*}
Plug Eq (\ref{eq:p(theta_k|X_k, I)}) in Eq (\ref{eq:lnp(X_k|I)}), we get
the approximation for marginal likelihood $\text{Pr}\left(\boldsymbol{X}_{k}\mid\boldsymbol{I}\right)$.

Though the approximation is consistent for any $\boldsymbol{\theta}_{k}^{*}$,
as pointed out by \citet{chib1995marginal}, the choice of $\boldsymbol{\theta}_{k}^{*}$
determines the efficiency of approximation. The approximation
is likely to be more precise with a $\boldsymbol{\theta}_{k}^{*}$
that is close to the true $\boldsymbol{\theta}_{k}$. A natural choice
for $\boldsymbol{\theta}_{k}^{*}$ is the posterior mean estimator
of $\boldsymbol{\theta}_{k}$ as calculated in Eq (\ref{eq:point_theta}).

\subsection{Point estimator for loss probabilities $\boldsymbol{\theta}$}

In the implementation of the Gibbs sampler, we integrate out the $\boldsymbol{\theta}$'s
from the model and run the collapsed Gibbs sampler, which improves the MCMC sampling efficiency. After obtaining the final
partitioning $\hat{\boldsymbol{I}}$, we want to calculate the point
estimators for the $\boldsymbol{\theta}$'s for the $K$ ECMs defined in
$\hat{\boldsymbol{I}}$, denoted by $\left\{ \hat{\boldsymbol{\theta}}_{1},\dots,\hat{\boldsymbol{\theta}}_{K}\right\} $.
For each ECM $k$, those branches with estimated high loss probabilities
$\hat{\theta}_{k,s}$ are evolutionary signature of ECM $k$ and distinguish
it from other ECMs. In Section \ref{sub:expansion}, we plug the estimated
parameters $\left\{ \hat{\boldsymbol{\theta}}_{1},\dots,\hat{\boldsymbol{\theta}}_{K}\right\} $
into the likelihood ratio statistics to identify novel genes that are not in
$\mathcal{G}$ but share close history with any of the $K$ ECMs.
The point estimator of $\theta_{k,s}$ is defined as the posterior
mean of $\theta_{k,s}$ conditional on $\mathbf{X}$ and $\hat{\boldsymbol{I}}$,
i.e. $\hat{\theta}_{k,s}\,=\,\mathbb{E}\left[\theta_{k,s}\mid\mathbf{X},\hat{\boldsymbol{I}}\right].$
To compute $\hat{\theta}_{k,s}$, we re-run the Gibbs sampler conditional
on $\hat{\boldsymbol{I}}$ to draw $M=1000$ samples $\boldsymbol{H}_{k}^{\left(1\right)},\dots,\boldsymbol{H}_{k}^{\left(M\right)}$
from $\text{Pr}\left(\boldsymbol{H}_{k}\mid\mathbf{X},\hat{\boldsymbol{I}}\right)$,
where $\boldsymbol{H}_{k}=\left\{ \boldsymbol{H}_{i}:\,I_{i}=k,i=1,\dots,n\right\} $.
$\hat{\theta}_{k,s}$ is approximated by the following Rao-Blackwellized
estimator \citep{liu1994covariance}:
\begin{equation}
\hat{\theta}_{k,s}
%\,=\,\sum_{\boldsymbol{H}_{k}}\mathbb{E}\left[\theta_{k,s}\mid\boldsymbol{H}_{k},\hat{\boldsymbol{I}}\right]\cdot\text{Pr}\left(\boldsymbol{H}_{k}\mid\mathbf{X},\hat{\boldsymbol{I}}\right)
\approx \frac{1}{M}\sum_{m=1}^{M}\mathbb{E}\left[\theta_{k,s}\mid\boldsymbol{H}_{k}^{\left(m\right)},\hat{\boldsymbol{I}}\right],\label{eq:point_theta}
\end{equation}
where
\begin{equation}
\mathbb{E}\left[\theta_{k,s}\mid\boldsymbol{H}_{k}^{\left(m\right)},\hat{\boldsymbol{I}}\right]=\frac{a+\sum_{i:\hat{I}_{i}=k}\mathbb{I}\left\{ H_{i,\delta\left(s\right)}^{\left(m\right)}=1,H_{i,s}^{\left(m\right)}=0\right\} }{a+b+\sum_{i:\hat{I}_{i}=k}\mathbb{I}\left\{ H_{i,\delta\left(s\right)}^{\left(m\right)}=1\right\} }.\label{eq:point_theta_pm}
\end{equation}

\section{Expansion step: identifying novel genes co-evolved with each ECM\label{sub:expansion}}

In the Partition step, CLIME 1.0 clusters the input set $\mathcal{G}$
into disjoint evolutionarily conserved modules (ECMs), simultaneously
inferring the number of ECMs and each gene\textquoteright s ECM membership.
The second step of CLIME 1.0, the Expansion step, identifies novel genes that are not in the input gene set $\mathcal{G}$ but share evolutionary history with any ECM $k$ identified in the
Partition step. The Expansion step is essential to CLIME 1.0 as the main goal
of it is to identify novel genes that are co-evolved with a subset
of $\mathcal{G}$. The underlying logic is that if a ECM $k$ consists
of a large number of genes of $\mathcal{G}$, then the other genes
not in $\mathcal{G}$ but share history with ECM $k$ are likely
functionally associated with $\mathcal{G}$. 

For each candidate gene $g$ and ECM $k$, $g=1,\dots,N$ and $k=1,\dots,K$,
we calculate the log-likelihood ratio (LLR),
\[
\text{LLR}_{g,k}\,=\,\log{\Pr}\left(\boldsymbol{X}_{g}\mid\hat{\boldsymbol{\theta}}_{k}\right)-\log\text{Pr}\left(\boldsymbol{X}_{g}\mid\hat{\boldsymbol{\theta}}_{0}\right),
\]
where the background null model $\hat{\boldsymbol{\theta}}_{0}$ is defined
as the estimated genome-wide average loss probabilities over all $N=20,834$
human genes. The estimation of $\hat{\boldsymbol{\theta}}_{0}$ is straightforward
and described in Section \ref{sub:gain_null_est}. In the LLR, the
first term $\log\text{Pr}(\boldsymbol{X}_{g}\mid\hat{\boldsymbol{\theta}}_{k})$
quantifies the likelihood that $\boldsymbol{X}_{g}$ was generated
from the HMM of ECM $k$, and the second term 
$\log\text{Pr}(\boldsymbol{X}_{g}\mid\hat{\boldsymbol{\theta}}_{0})$
quantifies the likelihood that $\boldsymbol{X}_{g}$ was generated
from the background null HMM. High value of $\text{LLR}_{g,k}$ indicates
that the HMM of ECM $k$ explains the phylogenetic profile $\boldsymbol{X}_{g}$
much better than the background null model, which suggests that gene
$g$ is more probable to share the same evolutionary history with
the genes in ECM $k$, than a randomly selected gene in human genome.

For each ECM, CLIME 1.0 scores all $N-n$ human genes, ranks them by LLR
scores, and reports the list of genes with LLR $>$ 0 (denoted by ECM+).
Compared to naïve metrics (e.g. Hamming distance, Pearson correlation
between phylogenetic profiles), this LLR statistic  measures co-evolution 
more appropriately and achieves substantially
higher prediction sensitivity and specificity (see Section \ref{sub:leave-one-out}).

\section{Pre-processing: estimation of gain branches $\boldsymbol{\lambda}$
and background null model $\boldsymbol{\theta}_{0}$\label{sub:gain_null_est}}

In the pre-processing stage, CLIME 1.0 infers the gain branch $\lambda_{i}$
for each gene $i$ and estimates the background null model parameter
$\hat{\boldsymbol{\theta}}_{0}$ for gene loss events from phylogenetic
profiles of all human genes in the input matrix. The null model is
an ECM-independent HMM whose branch-specific loss probabilities are
averaged over all genes in the human genome. 

We estimate $\boldsymbol{\theta}_{0}$ under the model that all $N=20,834$
human genes share the same loss probability vector $\boldsymbol{\theta}_{0}$,
i.e. $\boldsymbol{\theta}_{1}=\boldsymbol{\theta}_{2}=\cdots=\boldsymbol{\theta}_{N}=\boldsymbol{\theta}_{0}$,
and implement a Gibbs sampler to sample from the posterior distribution
of $\text{Pr}\left(\boldsymbol{\theta}_{0},\boldsymbol{\lambda}\mid\mathbf{X}_{1:N}\right)$.
We start the Gibbs sampler from the initial state with $\boldsymbol{\theta}_{0}=\left(0.03,\dots,0.03\right)$
and $\boldsymbol{\lambda}=\left(2S-1,\dots,2S-1\right)$. In each
step of the Gibbs sampler, we conduct the following steps: 
\begin{enumerate}
\item Draw $\lambda_{i}\sim\text{Pr}\left(\lambda_{i}\mid\boldsymbol{X}_{i},\boldsymbol{\theta}_{0}\right),\;i=1,\dots,N$.
\item Draw $\boldsymbol{H}_{i}\sim\text{Pr}\left(\boldsymbol{H}_{i}\mid\boldsymbol{X}_{i},\boldsymbol{\lambda},\boldsymbol{\theta}_{0}\right),\;i=1,\dots,N$
by the forward-backward procedure.
\item Draw $\boldsymbol{\theta}_{0}\sim\text{Pr}\left(\boldsymbol{\theta}_{0}\mid\mathbf{H}_{1:N},\boldsymbol{\lambda}\right),\;i=1,\dots,N$.
\end{enumerate}
Both conditional distributions $\text{Pr}\left(\lambda_{i}\mid\boldsymbol{X}_{i},\boldsymbol{\theta}_{0}\right)$
and $\text{Pr}\left(\boldsymbol{\theta}_{0}\mid\mathbf{H}_{1:N},\boldsymbol{\lambda}\right)$
are straightforward to sample from. $\text{Pr}\left(\lambda_{i}\mid\boldsymbol{X}_{i},\boldsymbol{\theta}_{0}\right)$
is a discrete distribution and for $s=1,\dots,2S-1$, 
\[
\text{Pr}\left(\lambda_{i}=s\mid\boldsymbol{X}_{i},\boldsymbol{\theta}_{0}\right)\,\propto\,\text{Pr}\left(\boldsymbol{X}_{i}\mid\lambda_{i}=s,\boldsymbol{\theta}_{0}\right)\text{Pr}\left(\lambda_{i}=s\right).
\]
We adopt a uniform prior on $\text{Pr}\left(\lambda_{i}=s\right)=1/\left(2S-1\right)$
and calculate likelihood function $\text{Pr}\left(\boldsymbol{X}_{i}\mid\lambda_{i}=s,\boldsymbol{\theta}_{0}\right)$
with dynamic programming outlined in Eq (\ref{eq:P(X|theta)}). $\text{Pr}\left(\boldsymbol{\theta}_{0}\mid\mathbf{H}_{1:N},\boldsymbol{\lambda}\right)$
is simply a product of Beta distributions, and each $\theta_{0,s}$,
$s=1,\dots,2S-2$, can be drawn independently. Similar to Eq (\ref{eq:point_theta}),
we define the point estimator of $\boldsymbol{\theta}_{0}$ as $\hat{\boldsymbol{\theta}}_{0}=\mathbb{E}\left[\boldsymbol{\theta}_{0}\mid\mathbf{X}_{1:N}\right]$
and approximate it with MCMC samples. Suppose we have $M$ MCMC
samples on $\boldsymbol{\lambda}$, denoted by $\boldsymbol{\lambda}^{\left(1\right)},\dots,\boldsymbol{\lambda}^{\left(M\right)}$.
For each gene $i=1,\dots,N$, we define $\hat{\lambda}_{i}$ as the {\it maximum a posteriori} (MAP) estimator approximated by MCMC samples,
\[
\hat{\lambda}_{i}\,\,=\,\,\underset{s}{\arg\max}\sum_{m=1}^{M}\mathbb{I}\left\{ \lambda_{i}^{\left(m\right)}=s\right\} .
\]

In both the Partition and the Expansion steps of CLIME 1.0, gain branches $\boldsymbol{\lambda}=\left(\lambda_{1},\dots,\lambda_{N}\right)$
are considered as known and fixed. An alternative way for estimating
the gain branch for each gene $i$ in $\mathcal{G}$ is to update
$\lambda_{i}$ in the Gibbs sampler of the Partitioning step and calculate
their posterior distributions. There are two reasons why we chose
to estimate the gain branch for each gene in the Pre-processing step and
kept it fixed in the later two steps. First, the gain branches can usually
be reliably estimated with little uncertainty. For example, if a gene
$i$ was truly gained at node $s$, then most likely we will observe
its presences only in $\boldsymbol{X}_{i}^{s}$, which informs us
that the gain event happened at node $s$. Second, by estimating the
gain branches at the Pre-processing step, we reduce the computation
complexity compared to a full model that updates $\boldsymbol{\lambda}$ at each MCMC iteration of the Partition step.

%%%%%%
\section{The extended model with uncertainty of phylogenetic tree}\label{sec:clime+}
\subsection{The extended model of CLIME 1.1}
Here we introduce the model of CLIME 1.1, which extends CLIME 1.0 by incorporating the uncertainty in phylogenetic trees. We keep the same notation as in the original CLIME 1.0. Conditioning on the tree structure $T$, we follow the same specification as in Eq (\ref{eq:CRP}). Additionally, we assume that the tree structure follows a prior $T\sim \mathcal{F_T}$, so  that jointly we have:
\begin{align*}
\boldsymbol{X}_i|\boldsymbol{H}^T_i, T &\sim P(\boldsymbol{X}|\boldsymbol{H}^T_{i}), \quad i=1,2,\dots n,\\
\boldsymbol{H}^T_{i}|\boldsymbol{\theta}_k^T, I_i=k, T & \sim P(\boldsymbol{H}^T_{i}|\boldsymbol{\theta}_{I_i}^T)\quad i=1,2,\dots n,\\
\boldsymbol{\theta}_k^T &\sim \prod_{s=1}^{2S-2} Beta(a,b), \quad k=1,2,\dots \\
I_i &\sim CRP (\alpha),  \quad i=1,2,\dots n,\\
T &\sim \mathcal{F}_T.
 \end{align*}

Here, the superscript $T$ indicates dependency on the tree structure, which will be suppressed in the following derivations for simplicity. In practice, we utilize the bootstrap samples or posterior draws of trees from the output of tree-constructing softwares to approximate the prior distribution $\mathcal{F}_T$. That is, suppose we have $N_T$ sampled tree structures $\{T_1, \dots, T_{N_T}\}$, we assume that $\mathcal{F}_T  = \frac{1}{N_T}\delta_{T_i}(T)$, where $\delta$ is the Dirac point mass.  This distribution is derived based on a probabilistic model of evolution and can well characterize the variability in the estimation of the evolutionary tree. 

\subsection{Posterior inference of CLIME 1.1 with Gibbs sampler}\label{CLIME+Gibbs}
We implement a collapsed Gibbs sampler \citep{liu1994collapsed} to draw from the posterior distribution, which cycles through the samplings of the hidden evolutionary history $\boldsymbol{H}$, the tree structure $T$, and the ECM label $\boldsymbol{I}$. The high-dimensional parameter vector $\boldsymbol{\theta}$ is integrated  out throughout the process similarly as what we did for CLIME 1.0 to improve the sampling efficiency.

\begin{enumerate}
\item Sampling $[\boldsymbol{H}\mid \boldsymbol{X}, \boldsymbol{I},T]$: 
For each gene $i$, we sample its evolutionary history $\boldsymbol{H}_i$ from  $\text{Pr}(\boldsymbol{H}_i | \boldsymbol{X}, \boldsymbol{H}_{-i}, T, \boldsymbol{I})$, which can be achieved by the same procedure described in  Section \ref{sub:H_sampling} to sample $\boldsymbol{H}_i$, conditioning on tree structure $T$.
\item Sampling $[\boldsymbol{I}\mid \boldsymbol{X}, \boldsymbol{H},T]$: 
For each gene $i$, we sample its cluster label $I_i$  from $\text{Pr}(I_i|\boldsymbol{I}_{-i}, \boldsymbol{X}, \boldsymbol{H}_{-i}, T)$, which, conditioning on tree structure $T$, can be similarly calculated as in Eq (\ref{eq:P(I=00003Dk|X,H,I)}).
\item Sampling $[T \mid \boldsymbol{X}, \boldsymbol{I}]$: We sample $T$ based on posterior 
$$
\text{Pr}(T|\boldsymbol{X}, \boldsymbol{I}) \propto \mathcal{F}_T(T) \text{Pr}(\boldsymbol{X}|T,\boldsymbol{I}).
$$
Since  the prior $\mathcal{F}_T $ is taken as the empirical distribution
$\frac{1}{N_T}\delta_{T_i}(T)$, we sample $T=T_i$ with probability proportional to $\text{Pr}(\boldsymbol{X}|T_i,\boldsymbol{I})$, where $\text{Pr}(\boldsymbol{X}|T_i,\boldsymbol{I})$ can be approximated by the method of \cite{chib1995marginal} as in Eq (\ref{eq:lnp(X_k|I)}). Note that the conditional distribution  $\text{Pr}(\boldsymbol{X}|T_i,\boldsymbol{I})$ will be used again in the Partition step for calculating $\arg\max_{\boldsymbol{I}}\text{Pr}(\boldsymbol{I}|\boldsymbol{X})$, and the Expansion step for calculating the LLR of novel genes.
\end{enumerate}

\subsection{Partition Step of CLIME 1.1}
We are mainly interested in estimating the ECM clustering labels of all input genes. Similar to CLIME 1.0, we adopt the MAP estimator $\hat{\boldsymbol{I}}=\arg\max_{\boldsymbol{I}}\text{Pr}(\boldsymbol{I}|X)$, approximated by searching through all MCMC samples of $\boldsymbol{I}$, i.e., 
\[
\hat{\boldsymbol{I}}\,=\,\underset{\boldsymbol{I}^{\left(m\right)}:\,m=1,\dots,M}{\arg\max}\: \text{Pr} \left(\boldsymbol{I}^{\left(m\right)}\mid\boldsymbol{X}\right).
\]
Specifically, 
\begin{align*}
\text{Pr}(\boldsymbol{I}|\boldsymbol{X}) \propto  \int \text{Pr}(\boldsymbol{X},\boldsymbol{I}|T)\mathcal{F}_T(T) dT
= \text{Pr}(\boldsymbol{I})\sum_{T_i} \frac{1}{N_T} \text{Pr}(\boldsymbol{X}|\boldsymbol{I},T_i),
\end{align*}
where the conditional distribution $\text{Pr}(\boldsymbol{X}|\boldsymbol{I},T_i)$ has been calculated in Step 3 of the Gibbs sampler in Section~\ref{CLIME+Gibbs}, and the prior $\text{Pr}(\boldsymbol{I})$ is assumed to be the Chinese restaurant process.

\subsection{Expansion step of CLIME 1.1}
Suppose a gene $g$'s phylogenetic profile being $\boldsymbol{X}_g$ ($g= 1, \dots, N$).  We calculate its  LLR for all ECMs, $k=1,\dots, K$, similarly as for CLIME 1.0, i.e.,
$$
LLR_{g,k} = \log \text{Pr}(\boldsymbol{X}_g|I_g=k, \boldsymbol{X}, \hat{\boldsymbol{I}}) -  \log \text{Pr}(\boldsymbol{X}_g|I_g=0, \boldsymbol{X}, \hat{\boldsymbol{I}}),
$$
where $I_g=0$ indicates the background null model.

We calculate the predictive likelihood by integrating out $\boldsymbol{\theta}_k$ and $T$:
 \begin{align*}
\text{Pr}(\boldsymbol{X}_g|I_g=k, \boldsymbol{X}, \hat{\boldsymbol{I}})
&=\int \text{Pr}(\boldsymbol{X}_g|I_g=k, \boldsymbol{\theta}_k, T)\text{Pr}(\boldsymbol{\theta}_k,T|\boldsymbol{X}, \hat{\boldsymbol{I}})dTd\boldsymbol{\theta}_k\\
%&\propto \int p(\boldsymbol{X}_g|I_g=k, \boldsymbol{\theta}_k, T, \hat{\boldsymbol{I}})p(\boldsymbol{X}, \boldsymbol{\theta}_k|T, \hat{\boldsymbol{I}})p(T)dTd\boldsymbol{\theta}_k\\
&= \int \text{Pr}(\boldsymbol{X}_g|I_g=k, \boldsymbol{\theta}_k, T)\text{Pr}(\boldsymbol{\theta}_k|T, \hat{\boldsymbol{I}}, \boldsymbol{X}) \text{Pr}(T | \boldsymbol{X}, \hat{\boldsymbol{I}}) dT d\boldsymbol{\theta}_k \\
&\propto\int \text{Pr}(\boldsymbol{X}_g|I_g=k, \boldsymbol{\theta}_k, T)\text{Pr}(\boldsymbol{\theta}_k|T, \hat{\boldsymbol{I}}, \boldsymbol{X})\text{Pr}(\boldsymbol{X}|T, \hat{\boldsymbol{I}}){\cal F}_T(T) dTd\boldsymbol{\theta}_k
\end{align*}
Note that  $\mathcal{F}_T  = \frac{1}{N_T}\delta_{T_i}(T)$, and % the posterior of $\boldsymbol{\theta}_k$ 
\begin{align*}
\text{Pr}(\boldsymbol{\theta}_k|T, \hat{\boldsymbol{I}}, \boldsymbol{X}) &= \int  \text{Pr}(\boldsymbol{\theta}_k|T, \hat{\boldsymbol{I}}, \boldsymbol{H}) \text{Pr}(\boldsymbol{H}|\boldsymbol{X}, T, \hat{\boldsymbol{I}}) d\boldsymbol{H},
\end{align*}
which can be approximated using the Gibbs sampling draws as
%where $p(\boldsymbol{H}|\boldsymbol{X}, T, \hat{\boldsymbol{I}})$ can be approximated by MCMC samples $\boldsymbol{H}^{(1)}, \dots, \boldsymbol{H}^{(M)}$, i.e.,
\begin{align*}
\text{Pr}(\boldsymbol{\theta}_k|T, \hat{\boldsymbol{I}}, \boldsymbol{X})  &\approx \frac{1}{M} \sum_{i=1}^M \text{Pr}(\boldsymbol{\theta}_k|T, \hat{\boldsymbol{I}}, \boldsymbol{H}^{(m)}).
\end{align*}
%By the linearity of likelihood as in {(\ref{eq:lnp(X_k|I)})}, 
Plugging in the foregoing integral, we have the following approximation
 \begin{align*}
\text{Pr}(\boldsymbol{X}_g|I_g=k, \boldsymbol{X}, \hat{\boldsymbol{I}})
%&=\int p(\boldsymbol{X}_g|I_g=k, \boldsymbol{\theta}_k, T, \hat{\boldsymbol{I}})p(\boldsymbol{\theta}_k|T, \hat{\boldsymbol{I}}, \boldsymbol{X})p(\boldsymbol{X}|T, \hat{\boldsymbol{I}})p(T)dTd\boldsymbol{\theta}_k\\
&\approx \frac{1}{N_T}\sum_{i=1}^{N_T} \left[\frac{1}{M}\sum_{m=1}^{M} \text{Pr}(\boldsymbol{X}_g|I_g=k, \bar{\boldsymbol{\theta}}_k^{(i,m)}, T_i, )\text{Pr}(\boldsymbol{X}|T_i,\hat{\boldsymbol{I}}) \right], 
\end{align*}
where $\bar{\boldsymbol{\theta}}_k^{(i,m)} = E(\boldsymbol{\theta}_k|T_i, \hat{\boldsymbol{I}},\boldsymbol{H}^{(m)})$ can be calculated by conjugate Beta distribution as in Eq (\ref{eq:point_theta_pm}); the predictive likelihood $\text{Pr}(\boldsymbol{X}_g|I_g=k, \bar{\boldsymbol{\theta}}_k^{(i,m)}, T_i)$ can then be calculated by dynamic programming introduced in Section \ref{sub:H_integration}; and 
%$$p(\boldsymbol{X} |T_i, \hat{\boldsymbol{I}} ) \approx \frac{1}{M} \sum_{i=1}^M p(\boldsymbol{X} |T_i, \hat{\boldsymbol{I}}, \boldsymbol{H}^{(m)} ).$$
the likelihood of input gene set $\text{Pr}(\boldsymbol{X}|T_i,\hat{\boldsymbol{I}})$ has been previously calculated in the step 3 of Gibbs sampler in Section~\ref{CLIME+Gibbs}.

\section{Simulation studies\label{sec:simstudy}}
We simulated the phylogenetic profile data from two models: a tree-based hidden Markov model and a tree-independent model where CLIME 1.0 and CLIME 1.1's model is mis-specified. The simulated input gene sets contain 50 genes, comprising a mixture of 5 ECMs, each with 10 genes, whose phylogenetic profiles were generated using the tree-based and tree-independent models. We analyzed the data with four methods: (1) CLIME 1.0; (2) CLIME 1.1;  (3) hierarchical clustering based on Hamming distance \citep{pellegrini1999assigning};(4) hierarchical clustering based on squared anti-correlation distance \citep{glazko2004detection}, where the distance between gene $i$ and $j$ is defined as $d_{i,j}=1-\left[\text{corr}\left(\boldsymbol{X}_{i},\boldsymbol{X}_{j}\right)\right]^{2}$.

For the tree-based hidden Markov model, we first used MrBayes \citep{ronquist2003mrbayes} to obtain 100  phylogenetic trees generated from the posterior distribution of the tree structure model based on 16 highly reserved proteins of 138 eukaryotic species \citep{bick2012evolutionary} and an additional prokaryote outgroup (139 species in total). For each simulation, we randomly picked one of the 100 tree structures, and generated the phylogenetic profiles and ECM assignments based on the tree-based HMM and this picked tree structure. Note that here we simulated uncertainties in the tree structure. Thus,
the original CLIME 1.0 with a single phylogenetic tree (the consensus) input runs the risk of tree misspecification for these simulated data.
%without taking the tree-building uncertainty into account. 
For each ECM, we first randomly selected one branch in the evolution tree to be the gain branch, and then, along its sub-tree, selected $N_{L}$ branches to be the potential gene loss branches and assign $P_{L}$ to be their gene loss probability to generate the phylogenetic profile of each gene. A higher $P_{L}$ leads to a more similar evolutionary history among the simulated genes in the same ECM, and a lower $P_{L}$ makes the underlying histories of genes less similar and adds more difficulty to the algorithms.
We simulated observation error with rate $q=0.02$, which is different from $q=0.01$ as pre-specified in CLIME 1.0 and CLIME 1.1's model. In addition, we simulated $N_{S}\in\left\{ 0,10,20,50\right\} $ singleton ECMs with one gene in each ECM as the background noise. Eventually, each input dataset contains a $(50+N_{s})\times 139$ binary matrix indicating the presence or absence of each gene in each species. 

For the tree-independent generating model in comparison, $N_{L}$ potential gene losses were randomly selected from 139 species without any reference to their evolutionary relations. Note that such a tree-independent model is equivalent
to a tree-based model when all the losses are constrained to happen exclusively on leaf branches. We range $ P_{L} \in \{0.6,0.7,0.8,0.9\}$ and $N_{L}\in \{4,6,8,10\}$ for both the tree-based model and the tree-independent model. Higher $N_{S}$ gives more noise and higher $P_{L}$ and $N_{L}$ indicate more independent loss events across various ECMs thus stronger signal.

\begin{figure}
\centering{}\includegraphics[width=\textwidth]{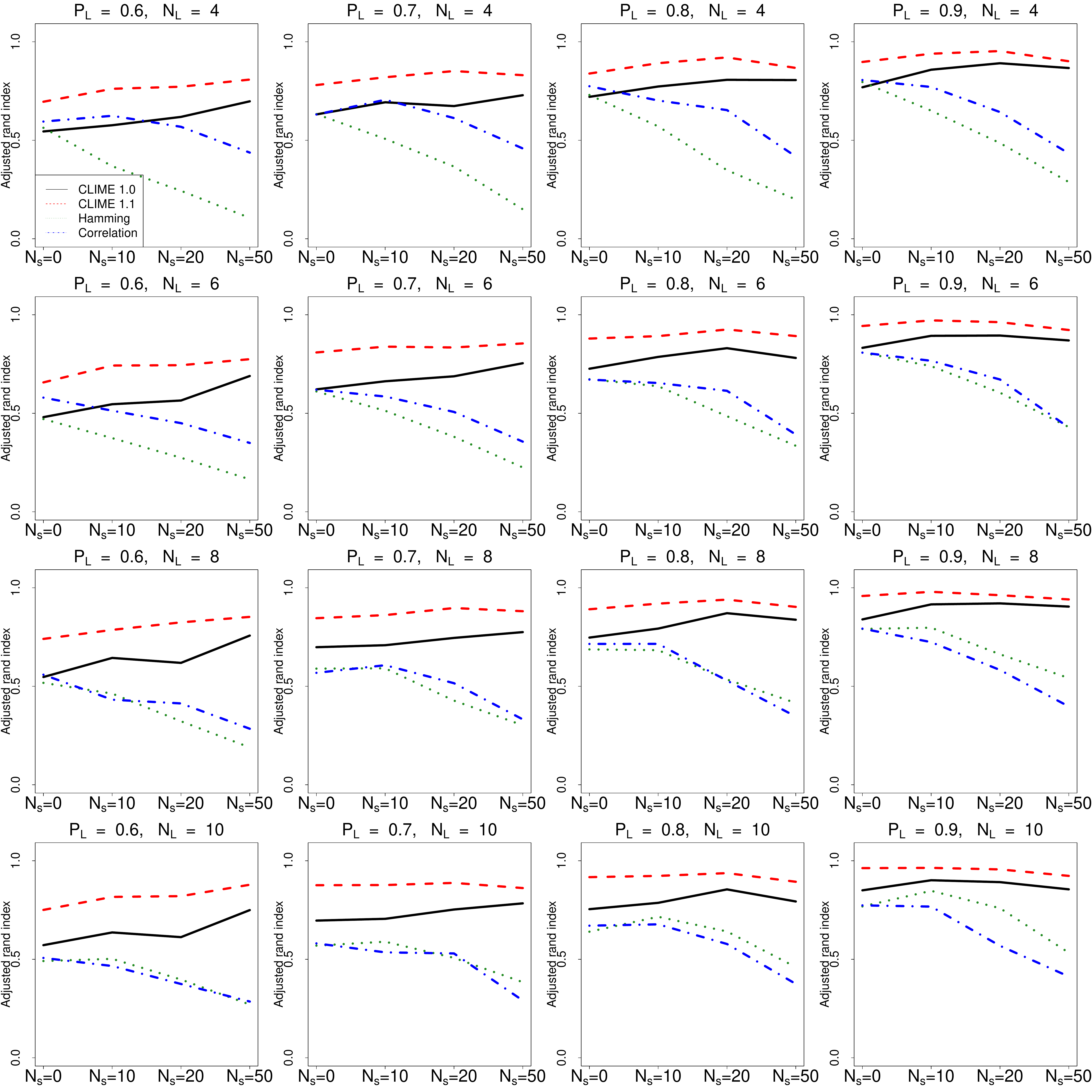}\caption{Simulation study results under tree-based model. Comparison of clustering accuracy (ARI) between CLIME 1.0 (black solid line), CLIME 1.1 (red dash),  hierarchical clustering by Hamming distance (green dot), and hierarchical clustering by anti-correlation (blue dot-dash). $N_{L}$: number of tree branches for each ECM to have non-zero loss probability. $P_{L}$: loss probability for the $N_{L}$ branches. $N_{S}$: number of singleton ECMs for each dataset.\label{fig:sim_results}}
\end{figure}

\begin{figure}
\centering{}\includegraphics[width=\textwidth]{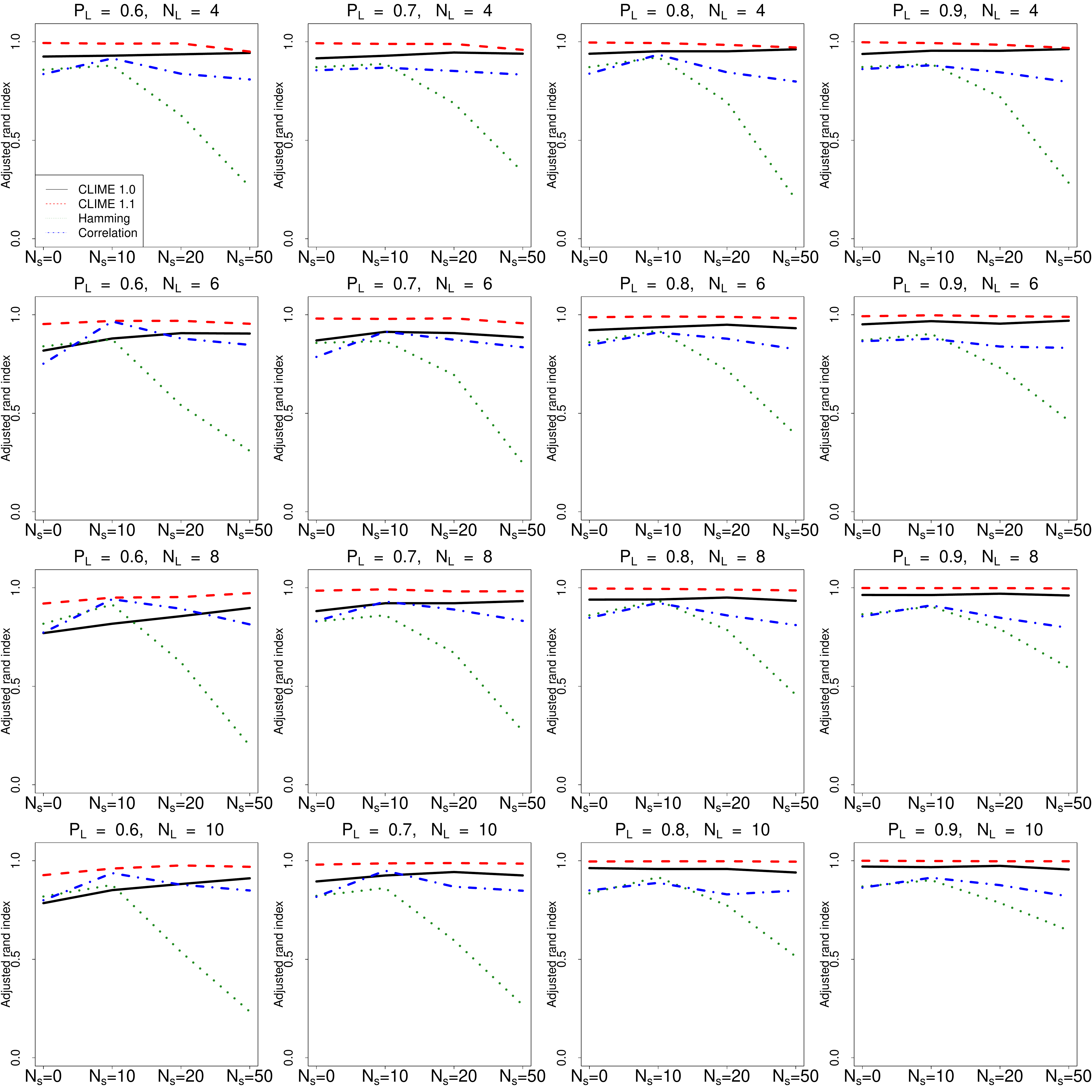}\caption{Simulation study results under tree-independent model. Comparison of clustering accuracy (ARI) between CLIME 1.0 (black solid line), CLIME 1.1 (red dash),  hierarchical clustering by Hamming distance (green dot), and hierarchical clustering by anti-correlation (blue dot-dash). $N_{L}$: number of tree branches for each ECM to have non-zero loss probability. $P_{L}$: loss probability for the $N_{L}$ branches. $N_{S}$: number of singleton ECMs for each dataset\label{fig:sim_results_nt}}
\end{figure}

\begin{figure}
\centering{}\includegraphics[scale=0.565]{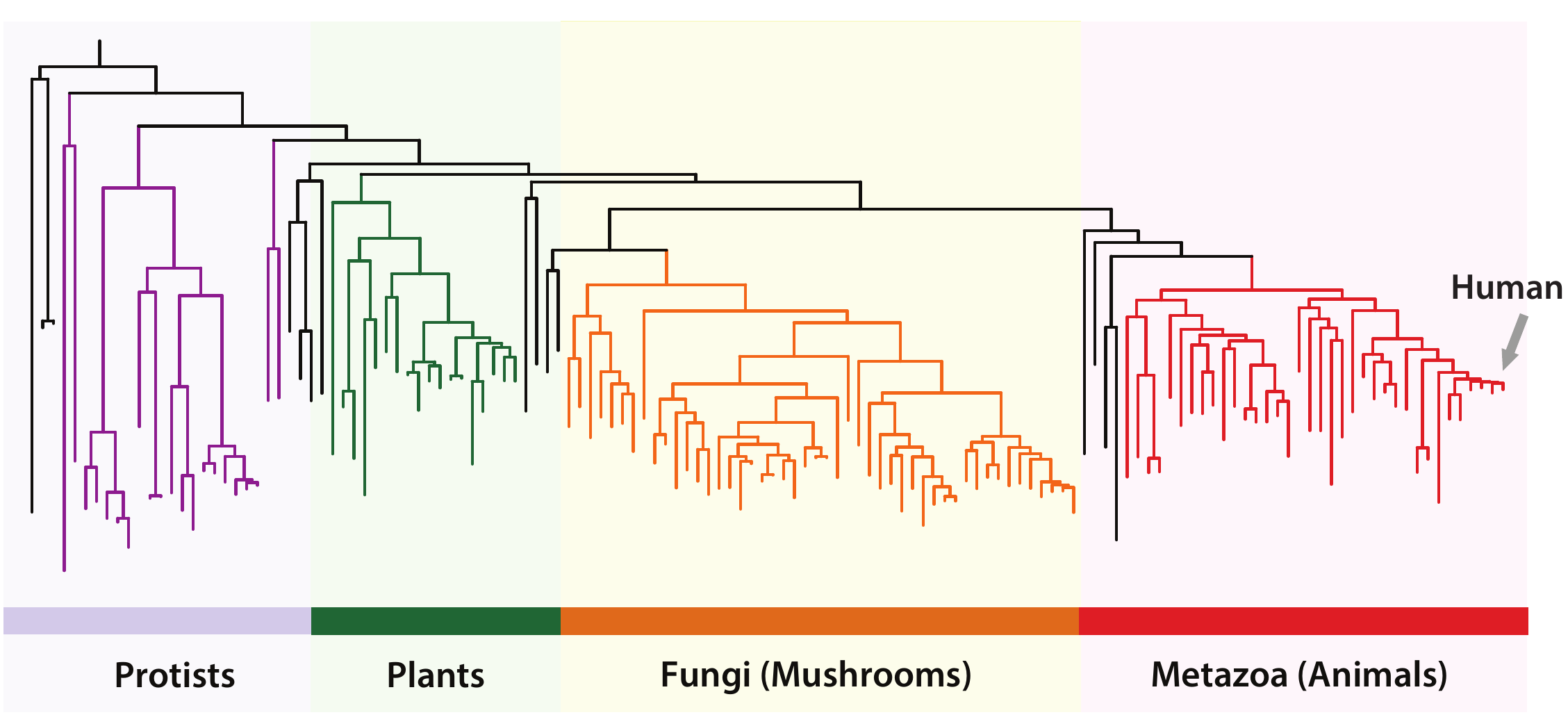}\caption{Phylogenetic tree in use with 138 eukaryotic species\textcolor{black}{{}
\citep{bick2012evolutionary}}. The tree consists of species in four
different eukaryotic kingdoms (Protists, Plants, Fungi and Animals),
labeled in four different colors. Human is the rightmost species on
the tree.\label{fig:evotree}}
\end{figure}

For each set of parameters, we simulated phylogenetic profile matrices for 20 times, applied all four methods, and adopted
the average adjusted Rand index (ARI) \citep{hubert1985comparing} between the estimated and true partitioning for these 20 simulated
datasets to evaluate clustering accuracy. For CLIME 1.0, to be consistent with the online software, we used the consensus phylogenetic tree built from 16 highly reserved proteins of 138 species \citep{bick2012evolutionary} with one outgroup prokaryote species as the single input tree structure, shown in Figure \ref{fig:evotree}. For CLIME 1.1, we included the 100 MrBayes samples described above as the input for empirical prior of the tree structure to account for estimation uncertainty. For hierarchical clustering, we used $10\%$ singleton genes as cutoff for clustering as adopted in \citep{glazko2004detection}. The complete simulation results for tree-based model and tree-independent model are reported in Figures \ref{fig:sim_results} and \ref{fig:sim_results_nt}, respectively.

As shown in Figure \ref{fig:sim_results}, when phylogenetic profiles were generated from a tree-based model of evolution with the risk of tree misspecification, CLIME 1.1 dominates all other clustering methods in terms of accuracy with the tree-uncertainty taken into account. CLIME 1.0, in general, also holds the lead over hierarchical clustering methods. The advantages of our tree-based Markov model are even more significant in scenarios where more loss events are present along the evolutionary history, i.e., more loss branches ($N_L \geq 6$) with higher ($P_L \geq 0.7$), to provide stronger signals for our tree-based model. Another feature of our methods is the robustness against the varying number of singleton ECMs, or the noise in clustering. As the noise level ($N_S$) increases, both CLIME 1.0 and CLIME 1.1 demonstrate consistent clustering accuracy, while hierarchical clustering methods show severe decay in performance. Notably, by incorporating the uncertainty of tree structure and weighting the clustering on the more probable tree structures, CLIME 1.1 further boosts the clustering accuracy of CLIME 1.0, where the latter draws inference based solely on a single possibly incorrect tree input. 

Simulations under the tree-independent model give all four methods a more even ground. Yet still, both CLIME 1.0 and CLIME 1.1 outperformed other benchmark methods in most of the simulation settings. Specifically, CLIME 1.1 maintained its domination over all other methods, sustaining the benefit of incorporating of tree structure viability. With a distribution of possible evolutionary trees to integrate, CLIME 1.1 takes advantage of the effect of model averaging through posterior updates of tree structure, and adapts more successfully to the change of the generative model. Both CLIME 1.0 and CLIME 1.1 maintained consistency in performance across varying simulation setting, while hierarchical methods, especially the one with Hamming distance, is very sensitive to the noise level ($N_S > 0$).

\section{Application to real data\label{sec:realdata}}

We next apply both CLIME 1.0 and CLIME 1.1 to several real datasets,
including two selected gene sets (mitochondrial complex I and proteinaceous extracellular matrix), as well as 409 manually
curated gene sets from OMIM (Online Mendelian Inheritance in Man)
\citep{hamosh2005online}, where each gene set consists of genes known
to be associated to a specific genetic disease. We show that
CLIME 1.0 and CLIME 1.1 enjoy advantages in clustering accuracy over existing
methods. Furthermore,
the results of clustering and expansion analysis by CLIME 1.0 and CLIME 1.1 on these
gene sets agree with established biological findings and also shed
lights on potential biological discovery on gene functions and pathway
compositions.

\subsection{Phylogenetic tree and matrix}

To facilitate the following analyses by CLIME 1.0, we used a single, consensus species tree that was published in \citep{bick2012evolutionary} consisting of 138 diverse, sequenced eukaryotes with an additional prokaryote outgroup. For the analyses by CLIME 1.1, we used 100 posterior samples obtained by MrBayes \citep{ronquist2003mrbayes} based on the 16 highly reserved proteins of 138 species used by \citep{bick2012evolutionary}. We used the phylogenetic profile matrix in \citep{li2014expansion} for all $N=20,834$ human genes across the 139 species. A greater diversity of the organisms in the input tree often leads to a greater power for CLIME 1.0 and CLIME 1.1, through the increased opportunity for independent loss events. 

\subsection{Leave-one-out cross validation\label{sub:leave-one-out}}

We compared CLIME 1.0 with Hamming distance and BayesTraits (BT) \citep{barker2005predicting, pagel2007bayestraits}, another phylogenetic-tree-based method for gene co-evolution analysis. We conducted leave-one-out cross-validation analysis on two selected pathway/gene sets (mitochondrial complex I and proteinaceous extracellular matrix) to evaluate the clustering accuracy of the three methods. Note that we here focus on the performance of CLIME 1.0, considering the computational demands of CLIME 1.1. In Section \ref{sec:complex1} and \ref{sec:OMIM}, we show that CLIME 1.0 and CLIME 1.1 give relatively consistent results in real pathway-based data analysis.

For each gene set, we applied CLIME 1.0 to all but one gene within a specific pathway as test set for ECM identification and then expand the identified ECMs with the remaining human genes ($\sim 20,000$ candidate genes). We varied the LLR threshold in the expansion step of CLIME 1.0 and repeated this leave-one-out procedure for all genes in the gene set to calculate the average sensitivity and specificity of the algorithm. Note that the true positive calls (sensitivity) are made when the left-out gene is included in the expansion list and false positive calls are made when genes outside the pathway are included in the expansion list of any established ECM. For comparison, we also conducted the same experiment with the Hamming distance method \citep{pellegrini1999assigning} and BayesTraits.

%\textcolor{blue}{
BayesTraits is computationally intensive as it evaluates genetic profiles in a pairwise manner (estimated $\sim 244$-hour CPU time for $44\times 20,000$ pairwise calculation, one leave-one-out experiment for a $44$-gene pathway test set; versus CLIME 1.0's $\sim 2$-hour CPU time). For efficiency in computation, we only subsampled $500$ genes from remaining ($\sim 20,000$) human genes as the candidate set for gene set expansion. We calculated the pairwise co-evolution p-values by BayesTraits between all genes in the leave-one-out test set and the candidate set, and made a positive call if the minimal p-value between the candidate gene and each gene in the test set is below certain threshold. Similarly, we varied the threshold to obtain the sensitivity and specificity of the algorithm.
%}

\begin{figure}[!h]
\centering
\includegraphics[width=0.9\textwidth]{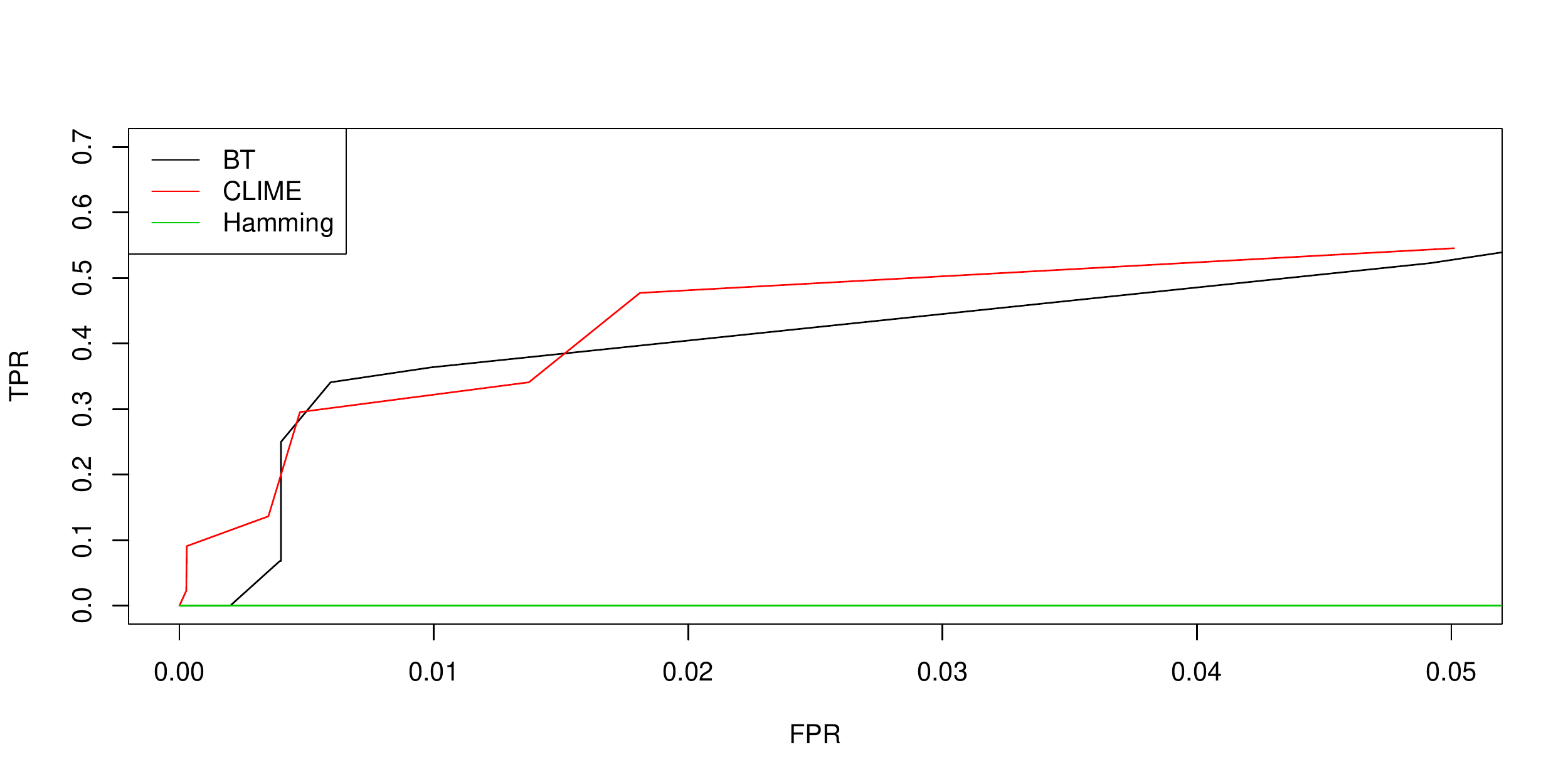}
\caption{Real data leave-one-out cross-validation on gene set: mitochondrial respiratory chain complex I.  Comparison of ROC curves
between CLIME 1.0, BayesTrait, and Hamming distance. \label{fig:loo_CI}}
\end{figure}

\begin{figure}[!h]
\centering
\includegraphics[width=0.9\textwidth]{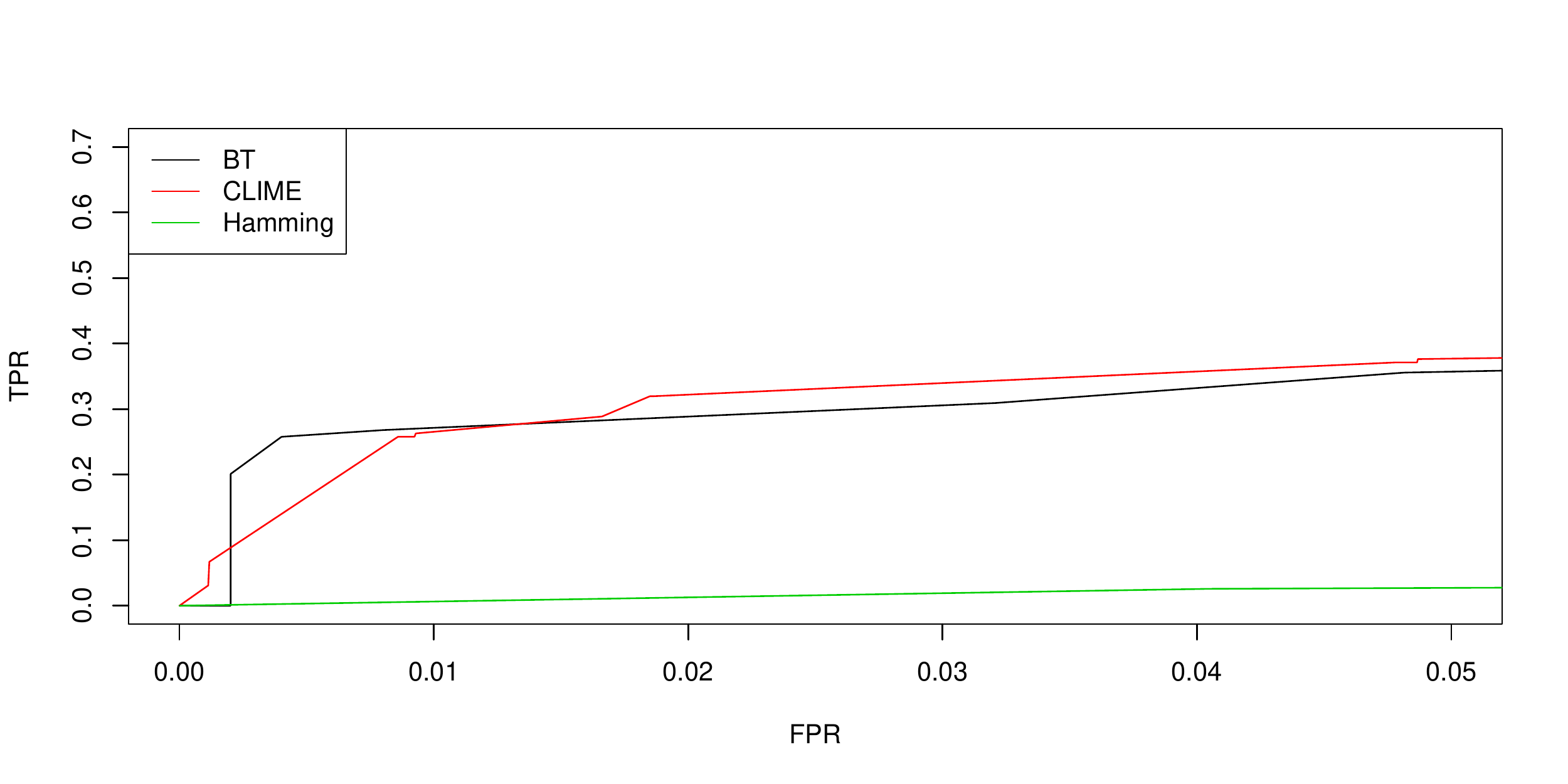}
\caption{Real data leave-one-out cross-validation on gene set: proteinaceous extracellular matrix.  Comparison of ROC curves
between CLIME 1.0, BayesTrait, and Hamming distance. \label{fig:loo_PEM}}
\end{figure}

We applied all these methods to two gene sets, mitochondrial respiratory chain complex I (44 genes), and proteinaceous extracellular matrix (194 genes) and report the receiver operating characteristic curves (ROC, true positive rate (TPR) versus false positive rate (FPR)) of all methods in Figure \ref{fig:loo_CI} and \ref{fig:loo_PEM} respectively.

Both CLIME 1.0 and BayesTraits dominated the Hamming-distance-based method, showing the strong advantage of incorporating the information from phylogenetic trees for functional pathway analysis based on genetic profiles. Compared with BayesTraits, CLIME 1.0 performed slightly better than BayesTraits in majority of the evaluation range of the ROC curve. Particularly, CLIME 1.0 dominated BayesTraits in both experiments when false positive rates  are under $0.2\%$, indicating CLIME 1.0's strength in providing accurate gene clustering with controlled mis-classification errors.

\subsection{Human complex I}\label{sec:complex1}
\begin{figure}
\centering

\includegraphics[width=\textwidth]{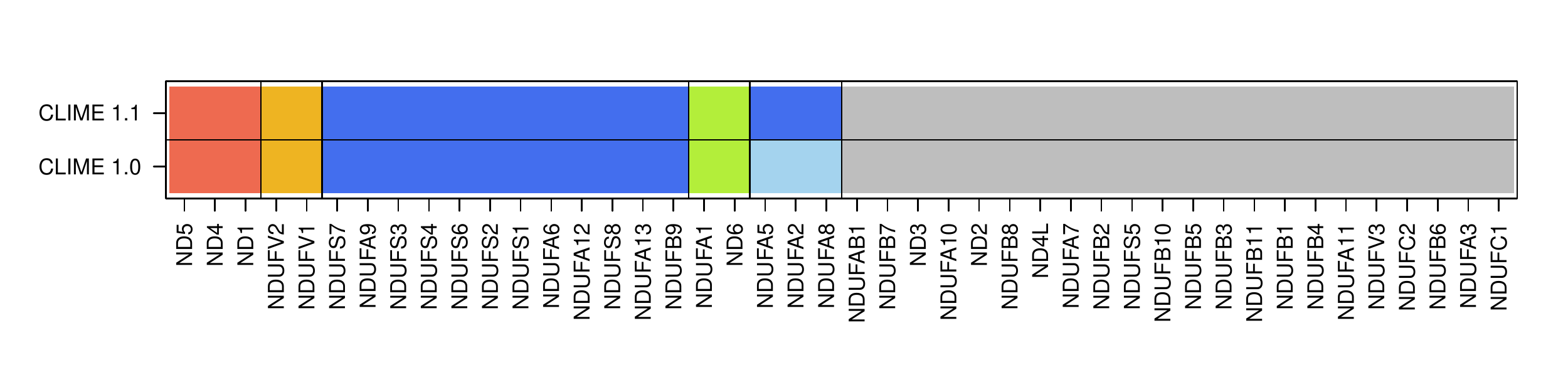}
\caption{Partition of 44 human Complex I genes by CLIME 1.0 and CLIME 1.1. Genes with same colored blocks are included in the same non-singleton ECMs. Grey color indicates singleton genes.}\label{fig:complex1}
\end{figure}

We compared CLIME 1.0 and CLIME 1.1 on a set of  44 human genes encoding complex I, the largest enzyme complex of the mitochondrial respiratory chain essential for the production of ATP \citep{balsa2012ndufa4}. CLIME 1.0 partitioned the 44 genes into five nonsingleton ECMs, and CLIME 1.1 gave nearly identical clustering (ARI:  0.962), as shown in Figure \ref{fig:complex1}, except that CLIME 1.1 combines the two ECMs by CLIME 1.0 that are related to nuclear DNA encoded subunits of the alpha subcomplex (with prefix NDUF)\citep{mimaki2012understanding}.  Both CLIME 1.0 and CLIME 1.1 identified an ECM containing only the subunits encoded by mitochondrial DNA (ECM1: ND1, ND4 and ND5, ECM strength by CLIME 1.0: $\phi=30.1$, CLIME 1.1: $\phi=30.1$), and an ECM comprising solely the core components of the N module in complex I (ECM2: NDUFV1 and NDUFV2, ECM strength by CLIME 1.0: $\phi=6.2$, CLIME 1.1: $\phi=6.7$)\citep{mimaki2012understanding}. A detailed report on the largest ECM (indexed ECM3, ECM strength by CLIME 1.0: $\phi=5.0$, CLIME 1.1: $\phi=5.8$) by both methods and their respective top extended gene sets (ECM3+) is shown in Table \ref{tab:complex1}. ECM3 mainly contains the nuclear-DNA-encoded subunits of complex I, including all four core subunits in the module Q of complex I (marked by asterisk). Among the top extended genes in  ECM3+, multiple complex I assembly factors and core subunits are identified (marked by boldface).

\begin{table}[ht]
{\centering
\setlength{\tabcolsep}{4pt}
\resizebox{\textwidth}{!}{
\begin{tabular}{r|llll||llll}
  \hline
    & \multicolumn{4}{c||}{CLIME 1.0}&  \multicolumn{4}{c}{CLIME 1.1} \\ 
      \hline
\multirow {4}{*} {ECM3}& NDUFS7* & NDUFA9 & NDUFS3* & NDUFS4 & NDUFS7* & NDUFA9 & NDUFS3* & NDUFS4 \\ 
   & NDUFS6 & NDUFS2* & NDUFS1 & NDUFA6 & NDUFS6 & NDUFS2* & NDUFS1 & NDUFA6 \\ 
   & NDUFA12 & NDUFS8* & NDUFA13 & NDUFB9 & NDUFA12 & NDUFS8* & NDUFA13 & NDUFB9 \\ 
   &\multicolumn{4}{c||}{}& NDUFA5 & NDUFA8 & NDUFA2 &\\
  \hline
  \hline
\multirow {5}{*} {ECM3+}& \textbf{NDUFAF5} & GAD1 & GADL1 & \textbf{NDUFAF7} & GAD1 & \textbf{NDUFAF7} & GADL1 & \textbf{NDUFAF5} \\ 
   & DDC & HDC & IVD & \textbf{NDUFAF6} & DDC & HDC & HSDL2 & CSAD \\ 
   & \textbf{NDUFV1} & ACADL & \textbf{NDUFV2} & CSAD & \textbf{NDUFAF1} & CPSF6 & IVD & GAD2 \\ 
   & \textbf{NDUFAF1} & CPSF6 & GAD2 & HSDL2 & ACADL & \textbf{NDUFAF6} & HPDL & HPD \\ 
   & RHBDL1 & MCCC2 & HPDL & ACADVL & \textbf{NDUFV1} & \textbf{NDUFV2} & RHBDL1 & MCCC2 \\ 
   \hline
\end{tabular}}
\label{tab:complex1}}
\caption{ECM3 and its extension  ECM3+ derived from the set of 44 human Complex I genes by CLIME 1.0 and CLIME 1.1. Asterisk indicates core subunits of complex I; boldface indicates predictions with recent experimental supports for functional association with the input set.}
\end{table}

\subsection{Gene sets related to human genetic diseases}\label{sec:OMIM}

\begin{figure}
\centering{}\includegraphics[scale=0.72]{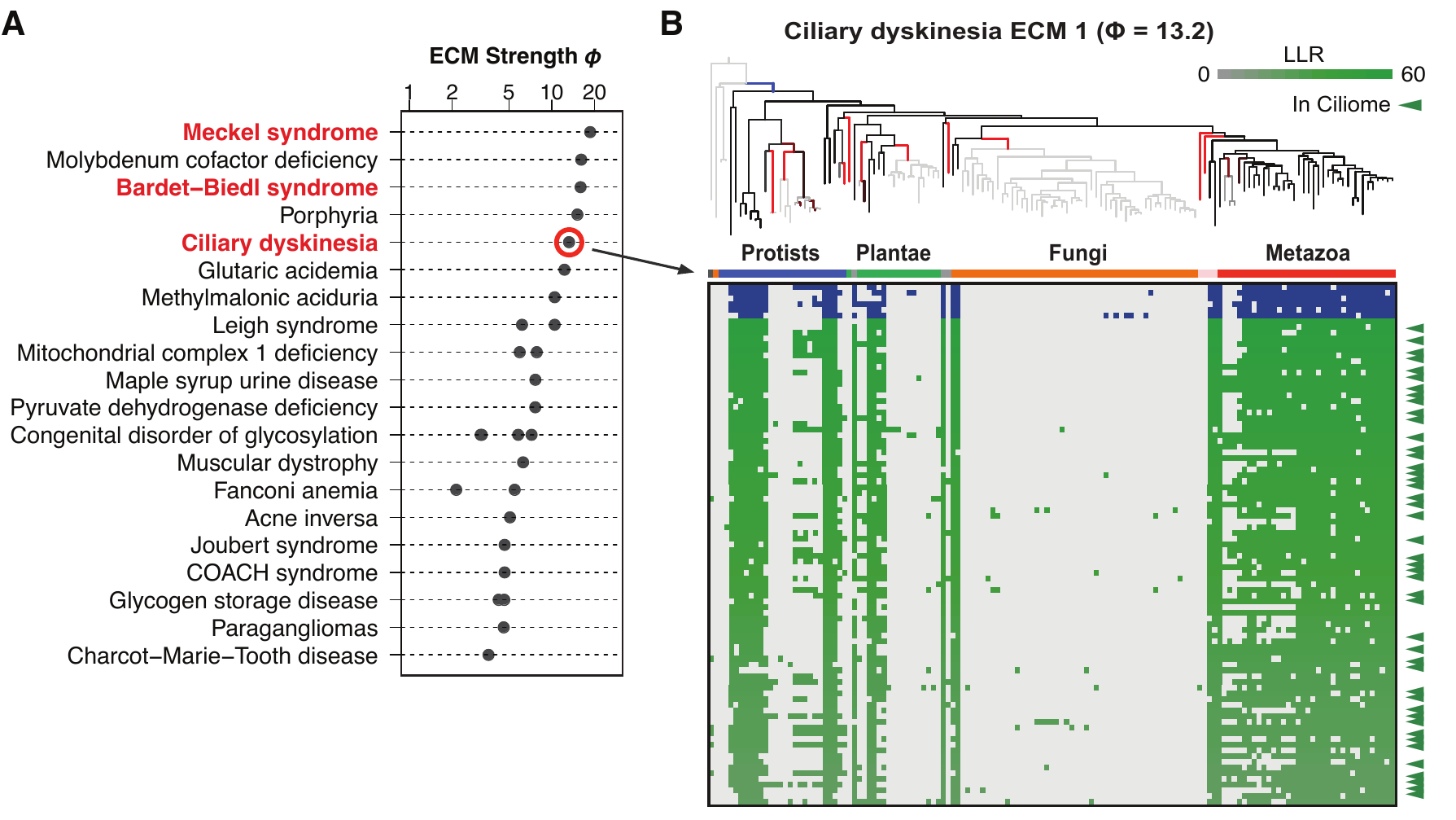}\caption{(A) Top 20 OMIM gene sets with highly informative ECMs by CLIME 1.0, ranked by
strength of the top ECM. All non-singleton ECMs are shown as separate
dots. Three gene sets related to human ciliary dysfunction are highlighted
in red. (B) ECM 1 for ciliary dyskinesia gene set. The inferred gain/loss
events are indicated by blue and red tree branches. Blue/white and
green/white matrices show phylogenetic profiles of ECM and expanded
genes, respectively. Green arrows indicate predicted new genes that
are supported by Ciliome database. \label{fig:omim}}
\end{figure}

We performed the analysis by CLIME 1.0 on 409 manually
curated gene sets from OMIM (Online Mendelian Inheritance in Man)\citep{hamosh2005online}, where each gene set consists of genes known
to be associated with a specific genetic disease. CLIME 1.0 identified non-singleton ECMs in 52 of these 409 gene sets (check \url{http://www.people.fas.harvard.edu/~junliu/CLIME/} for complete results). Figure~\ref{fig:omim} shows the top 20 disease-associated
gene sets with the highest strength ECMs. For gene sets related to diseases such as Leigh syndrome, mitochondrial complex I deficiency, and congenital disorder of glycosylation, multiple high-strength ECMs were identified by CLIME 1.0, which suggests that functionally distinct sub-groups may exist in these gene sets. We note that among top five gene sets, three are related to the human ciliary disease (highlighted in red). Specifically, the only non-singleton ECM ($\phi=13.2$)
for ciliary dyskineasia, defined by having more than 15 independent loss events,
is fully displayed in Figure \ref{fig:omim}B. The expansion list contains 100
novel genes with $\text{LLR}>0$. As illustrated by the heat map in Figure \ref{fig:omim}B, all genes in the ciliary dyskineasia ECM and its expansion list share a remarkable consensus in their phylogentic profiles.
%, which indicates the success of gene clustering by CLIME. 
Furthermore, about 50 of the 100 expansion genes belong to the Ciliome database \citep{inglis2006piecing}, an aggregation of data from seven large-scale experimental and computational studies, showing strong functional relevance of CLIME 1.0's expansion prediction.  
%The results demonstrate that CLIME can provide new perspectives on known functional gene sets, especially in the evolutionary context, and insights into the functionality of those less-studied genes.}

We next compared CLIME 1.1 with CLIME 1.0 on this ciliary dyskinesia gene set. The ECM partition by CLIME 1.1 is identical to CLIME 1.0, providing a strong support of such a subgroup structure among the ciliary-dyskinesia-related genes. We further compared the extended gene sets (ECM+) obtained by CLIME 1.0 and CLIME 1.1. Among the top 100 predicted genes, 89 are shared by CLIME 1.0 and CLIME 1.1, with top 20 reported in Table \ref{tab:OMIM}. Majority of the new members predicted by CLIME 1.0 and CLIME 1.1 can be validated as having functional association with cilia (cross-referenced by GeneCards: \url{https://www.genenames.org/}). In addition, the top four predicted genes have been found related to the primary ciliary dyskinesia \citep{horani2016genetics}, further demonstrating the promising power of CLIME 1.0 and CLIME 1.1 in the prediction of functional relevance.

%\subsection{Parameter estimate}

\begin{table}[ht]
\centering
\scriptsize
\setlength{\tabcolsep}{4pt}
\resizebox{0.9\textwidth}{!}{
\begin{tabular}{r|llll||llll}
  \hline
    & \multicolumn{4}{c||}{CLIME 1.0}&  \multicolumn{4}{c}{CLIME 1.1} \\ 
  \hline
\multirow {2}{*} {ECM} & RSPH4A & HEATR2 & RSPH9 & CCDC39 & RSPH4A & HEATR2 & RSPH9 & CCDC39 \\ 
   & CCDC40 & DNAAF2 &  &  & CCDC40 & DNAAF2 &  &  \\ 
     \hline
       \hline
\multirow {4}{*} {ECM+} & \textbf{RSPH6A}* & \textbf{CCDC65}* &  \textbf{RSPH3}* & \textbf{C6orf165}* & \textbf{RSPH6A}* & \textbf{CCDC65}* & \textbf{C6orf165}* & \textbf{RSPH3}* \\ 
 &  \textbf{DRC1}* & \textbf{SPEF1} & \textbf{PIBF1} & SPATA4 & \textbf{CCDC113} & \textbf{DRC1}* & \textbf{SPEF1} & \textbf{PIH1D3}* \\ 
   & \textbf{MAATS1} & \textbf{CCDC113} & \textbf{CCDC147} & ODF3 & SPATA4 & \textbf{MAATS1} & \textbf{CCDC147} & \textbf{PIBF1} \\ 
   & \textbf{C21orf59}* & \textbf{SPAG16} & \textbf{IQUB} & RIBC2 & ODF3 &\textbf{ IQUB} & \textbf{CCDC135} & CCDC146 \\ 
   & CCDC146 & \textbf{CCDC135} & \textbf{CCDC63} & \textbf{PIH1D3}* & \textbf{TTC26} & \textbf{SPAG16} & \textbf{CEP164} & \textbf{CCDC13} \\ 
   \hline
\end{tabular}}
\caption{The nonsingleton ECM and its extension ECM+ of the ciliary dyskinesia gene set by CLIME 1.0 and CLIME 1.1. For ECM+, boldface indicates predictions for functional association with the input set; asterisk indicates direct association with ciliary dyskinesia disease by recent experimental or human genetic support.}\label{tab:OMIM}
\end{table}

\section{Discussion\label{sec:discussion}}

%We previously introduced  CLIME  for partitioning and expanding functional gene sets based on inferred models of shared evolution  \citep{li2014expansion}. Here we provide the full Bayesian model and computational details of CLIME, perform new simulation studies, and apply it to gene sets related to Mendelian diseases. 
 Instead of integrating the pairwise co-evolution information 
of the genes in the input gene set in an {\it ad hoc} way,  CLIME 1.0 explicitly models multiple genes in a functional gene set as a set of disjoint gene modules, each with its own evolutionary history.
Leveraging information from multiple genes and modeling profile errors
are critical because phylogenetic profiles are often noisy due to
incomplete assemblies/annotations and errors in detecting distant
homologs.  Furthermore, CLIME 1.0 automatically infers the number of modules and gene assignments to each module. As an extension, CLIME 1.1 inherits these strengths of CLIME 1.0 and enhances its robustness and accuracy by incorporating uncertainty of evolutionary trees. CLIME 1.1, thereby, takes into account the estimation error in the tree estimation process, as well as the variability of phylogenetic relationships among genes.
Simulation studies and leave-one-out cross-validations on real data showed that CLIME 1.0 achieved a significantly improved accuracy in detecting shared evolution compared with benchmark methods we tested. CLIME 1.1 further adds to CLIME 1.0 with improved robustness and precision. 

Applications of CLIME 1.0 and CLIME 1.1 to real data testified the algorithms' excellent power in predicting  functional association between genes and in providing guidance for further biological studies (see \citep{li2014expansion} for more details).{
Based on our exemplary pathway/gene set data, CLIME 1.0 and 1.1 showed a great consistency in identifying evolutionarily conserved subsets of genes, and demonstrated high accuracy in recovering and predicting functionally connected gene groups. CLIME 1.1 further added in with discoveries of improved robustness and relevance.}

%We believe that CLIME 1.0 and 1.1 can serve as powerful tools to aid biologists in providing insights on genes' evolutionary and functional connectivities, and to complete the picture of biological pathways among human genes.}

{Specifically, in our investigations of the 44 complex-I-encoding genes, both CLIME 1.0 and 1.1 were able to identify subgroups of genes encoding different functional modules of complex I, and connect assembly factors associated with each module. CLIME 1.1 added to CLIME 1.0 by combining the two subgroups with nuclear DNA encoded subunits, further improving the biological interpretation of the clusters. This helps provide insights on the complete picture of complex I's catalyzing process and mechanism. We also applied our methods to more than 400 gene sets related to human genetic diseases, where CLIME 1.0 and 1.1 showed great potentials in predicting genes' functional associations with human genetic diseases. Focusing on the ciliary dyskineasia, both CLIME methods established novel connections between classic disease-driven genes and other cilia-related genes from the human genome. CLIME 1.1 furthered  prediction relevance with 5\% more cilia-related genes among the top predictions. Most notably, the top four predicted genes by both CLIME methods have been validated by recent studies on primary ciliary dyskineasia. This prompts biologists with a great confidence in using CLIME as a powerful tool and 
in following up CLIME's findings for further experimental validations and studies on such human genetic diseases.}

To trade for a gain in predication accuracy, CLIME 1.0 demands a comparatively high computational capacity. The computational complexity is about $O(Sn^{2})$
per MCMC iteration in the Partition step. For CLIME 1.1, with incorporation of tree uncertainty, the step-wise computational complexity is about $O(N_TSn^2)$. {In practice, to ensure computational efficiency, we recommend implementing CLIME 1.0 firstly for a general, large-scale exploration and CLIME 1.1 for more focused, follow-up analyses and validations, as demonstrated in the Section \ref{sec:complex1} and \ref{sec:OMIM}.}

As shown in simulation studies, CLIME 1.0 and CLIME 1.1 gain most of its prediction
power from the abundance of independent gene loss events through the evolutionary
process. In fact, independent gene losses create variability of phylogenetic
profiles between distinctive gene clusters, thus providing a strong
signal for CLIME 1.0 and 1.1 to make inference on. Similarly, in real data we
observe that CLIME 1.0 and 1.1\textquoteright s power is derived from the diversity
of species genomes, as it provides us opportunity to observe more
shared loss events. In recent years, the availability of completely
sequenced eukaryotic genomes is dramatically increasing. With growing
abundance and quality of eukaryotic genome sequences, the power of
CLIME 1.0 and 1.1 will increase as evolutionarily distant species are more likely
to possess abundant gene loss events, and thus stronger signals for CLIME 1.0 and 1.1
to extract. 

Further improvements of the model are possible. Currently, we do not estimate $q$ but set it as $0.01$ based on our prior knowledge on the observation error rate. Though we observe that the model is robust to $q$, it is more statistically rigorous to estimate $q$ from data. Furthermore, as there is variation between the quality of sequenced genomes, we can further assume that different species have different
mis-observation rates with independent priors, which can be estimated through posterior updating. Admittedly, point estimates for cluster labels by MAP provide an interpretable representation of the posterior results, especially convenient for scientists to conduct follow-up analysis or experiment. We may also consider alternative representation of the posterior on the cluster assignment, for example, the co-assignment probability for genes.

%$q_{s}$, $s=1,\dots,S$. Assigning independent Beta priors on $q_{1},\dots q_{S}$, it is straightforward to compute conjugate Beta conditional distributions $p\left(q_{s}\mid\mathbf{X},\boldsymbol{H}\right)$,$s=1,\dots,S$, and draw $q_{1},\dots,q_{S}$ samples in the Gibbs sampler. 

The results, a C++ software implementing the proposed method, and an online
analysis portal are freely available at \url{http://www.gene-clime.org}.
The website was previously introduced in \citep{li2014expansion}.

\section*{Acknowledgment}

This research was supported in part by the NSF Grant DMS-1613035, NIGMS Grant R01GM122080, and NIH Grant R35 GM122455-02. VKM is an Investigator of the Howard Hughes Medical Institute. 

\bibliographystyle{imsart-nameyear}
\bibliography{clime}

\end{document}